\documentclass{article}[11pt]
\usepackage{graphicx}
\usepackage{fullpage}

\RequirePackage{amsthm,amsmath,amsfonts,amssymb}
\RequirePackage[authoryear]{natbib}
\RequirePackage[colorlinks,citecolor=blue,urlcolor=blue]{hyperref}
\RequirePackage{graphicx}

\usepackage{enumitem}

\newcommand{\mE}{\ensuremath{\mathbb E}}
\newcommand{\mP}{\ensuremath{\mathbb P}}

\newcommand{\ind}{\perp \!\!\! \perp}

\newtheorem{algo}{Procedure}[section]

\theoremstyle{plain}

\newtheorem{theorem}{Theorem}[section]

\theoremstyle{remark}
\newtheorem{definition}[theorem]{Definition}
\newtheorem{example}{Example}[section]

\newtheorem{remark}{Remark}[section]

\begin{document}

\title{Replicability Across Multiple Studies}

 \author{
    Marina Bogomolov\\ Faculty of Data and Decision Sciences, 
Technion - Israel Institute of Technology, Haifa, Israel
    \\{marinabo@technion.ac.il}\\
    Ruth Heller\hspace{.2cm}\\Department of Statistics and Operations Research, Tel-Aviv Universit, Tel-Aviv, Israel\\{ruheller@gmail.com}
}    
\maketitle

\begin{abstract}
Meta-analysis is routinely performed in many scientific disciplines.  This analysis is attractive since discoveries are possible even when all the individual studies are underpowered.   However, the meta-analytic discoveries may be entirely driven by signal in a single study, and thus non-replicable.  
Although the great majority of meta-analyses carried out to date do not infer on the replicability of their findings, it is possible to do so.  We provide a selective overview of analyses that can be carried out towards establishing replicability of the scientific findings. We describe methods for the setting where a single outcome is examined in multiple studies (as is common in systematic reviews of medical interventions), as well as for the setting where multiple studies each examine multiple features (as in genomics applications).  We  also discuss some of the current shortcomings and future directions.
\end{abstract}

\section{Introduction}

Meta-analysis is routinely performed in many scientific disciplines. In the health sciences, it is common to  rely on systematic reviews and meta-analyses in order to inform practitioners of the best standard of care. For example, the Cochrane Collaboration carries out systematic reviews of health care interventions \citep{Higgins22}.  In large-scale analyses it is common to combine  several independent studies that examine the same outcome in order to increase the discovery power. For example, in genome wide association studies (GWAS), every study examines many genotypes, and by combining several studies that examine the same phenotype it is typically possible to discover more associations between genotype and phenotype than in an individual  study (see, e.g., \citealt{franke2010genome}).
The great majority of meta-analyses carried out to date do not infer on the replicability of their findings. Realizing that inference towards replicability of findings is possible from the  studies meta-analyzed \citep{BHY09}, there are more and more works that target replicability from the available set of studies for meta-analysis.

Lack of replicability of scientific findings has been of great concern in the last two decades. In an influential paper, \cite{Ioannidis05}  claimed that the fraction of false positive findings in many domains in science is  too high.
The false positives in the published research may occur due to  publication bias, or selective reporting, or other questionable research practices that inflate reported effect sizes, or mere bad luck. 
A common practice for corroborating findings is to bring forth  prior findings published that support the novel findings. This practice assumes that there are no false positives in the published research and may therefore be problematic, potentially causing proliferation of false findings as `scientific' facts.  
In order to reduce the lack of field-wise replicability, there was  a suggestion to lower the standard level for publication as  a discovery, $\alpha$, from 0.05 to 0.005. This suggestion was  supported by 72 researchers that are authors of the paper by \cite{Benjamin18},
  who believe that a leading
cause for published false  discoveries is that the statistical
standards of evidence for claiming new
discoveries in many fields of science are
 too low. A clear consequence of application of their suggestion is great power loss to make true discoveries. See Remark \ref{rem-Benjamin} for a further discussion of this suggestion.

An alternative way of raising the standard of evidence is to aim the analysis at replicable, rather than single, discoveries. For example, in GWAS, the demand for establishing replicability was often enforced by funders and journals, and GWAS findings are highly replicable \citep{Marigorta18}. 
Suppose we have at our disposal two or more studies that examine the same (or a similar) problem. These  may be a primary and then a follow-up study, or  a group of studies that are meta-analyzed together. 
 {\it Replicability analysis} refers to an analysis carried out towards identification of the  replicable scientific findings.  
  The following two observations regarding {\it replicability analysis}  are important:   
\begin{itemize}
    \item It is  more challenging than the typical analysis carried out in order to identify that a signal is present in at least one study,
    since replicability analysis takes care to rule out the possibility that the finding is  entirely driven by a single study. 
    \item It can always be carried out along with a meta-analysis, adding to the meta-analytic findings a quantitative measure of strength of evidence towards replicability of the findings.  
\end{itemize}

When  repeating a study, each repeated study  necessarily differs from the original study \citep{Nosek22}.  Therefore, the efforts to corroborate true scientific findings may fail  because  these findings are very specific to the study, i.e., they cannot be generalized to the conditions of a repeated study (which differs from the original study  in many ways). Even if the innumerable differences from the original study are believed to be irrelevant for obtaining the evidence about these true scientific findings,  the power may not be high enough in order to corroborate the findings. Recognizing the statistical challenge of potentially low power, the statistical community has developed methods specifically aimed at  discovering the replicable scientific findings (i.e., the findings that are corroborated by multiple studies). We shall provide our selective review of these methods. They can be useful for practitioners, and if indeed used often then we expect that they will lead to a reduction in the field-wise false positive findings.

After discussing the definition of replicability and introducing our notation, we review separately the case of a single feature (in \S~\ref{single}), and of  multiple features (in \S~\ref{sec:frequentist}  and \S~\ref{sec:Bayesian}). 
Can replicability be established if only a single study is available? Controlling for inflation of false positives in the single study is the first important step towards replicable research, and there is an abundance of multiple comparisons methods for this purpose. 
However, even with a  severe multiple testing correction, one can obtain findings that will not be replicable in other studies. This may happen, for example, because of  hidden biases. Of course, it is not possible to establish replicability by splitting the sample at random and thus creating multiple `studies' examining the same problem, because the same biases will be present in all these `studies.'
However, if it is possible to split the data  into subgroups based on certain important covariates (e.g. age, gender, ethnicity),
establishing replicability is possible even within a single study. This setting is addressed in  \S~\ref{sec:single study}. 

\section{The definition of replicability of scientific findings}\label{sec-definigreplicability}

Since studies necessarily differ, we should not expect the same results across studies. We should not even expect the underlying (unknown) effects to be the same, since we cannot rule out the possibility that the sampling and measurement variations across studies had an effect on the underlying parameters of interest. 

There is no consensus on how to quantify or measure replicability.  Various definitions of replicability, and no-replicability, were considered in  \cite{Nosek22,Fithian-AoAS-2020, Mathur19, Patil19,Hedges18, Goodman16, Simonsohn15}  and references within. It is necessary to define what replicability actually is in order to be able to assess the extent of the evidence towards replicability of findings. The answer should be in terms of relevant parameters, which are the underlying quantities that relevant summary statistics aim to estimate. For example, for a single feature (or outcome) in two studies: a reasonable definition is that the null hypothesis is false in both studies (with both parameters pointing in the same direction); a relevant summary statistic is the maximum  $p$-value; and a natural approach is to declare a finding as replicable if the maximum of the two $p$-values is at most $\alpha$.  Definitions of no-replicability include: 
 non-negligible statistical heterogeneity of effects;   mixed effect directions across studies;  an effect decline from an original study to a follow-up study. 

\subsection{$r$ out of $n$ replicability}
We consider the setting where $m$ features are examined in $n\geq 2$ independent studies. Let $\theta_{ij}$ denote the true parameter for feature $i$ in study $j$. In GWAS,  the parameter $\theta_{ij}$ may be a measure of association of the $i$th genotype  with the phenotype  in the  $j$th study. In a systematic review that examines the effects of a health care intervention,  $\theta_{ij}$ is the average treatment effect on the $i$th outcome in the $j$th study.

The research question regarding feature $i$ in study $j$  is
addressed by testing the null hypothesis $H_{ij}: \theta_{ij}\in \Theta^0_{ij},$ for $i=1, \ldots, m,$ $j=1, \ldots, n,$ where  $\Theta^0_{ij}$ is a certain set of possible values of $\theta_{ij}.$
If the alternative is right sided, then
$H_{ij}: \theta_{ij}\leq \theta^0_{ij}$, for a certain constant $\theta^0_{ij},$ and $H_{ij}$ is false if  $\theta_{ij}>\theta_{ij}^0$. Let $n_i^+\in \{0,\ldots,n \}$  be the number of parameters for feature $i$ for which $H_{ij}$ is false in this case, i.e. $n_i^+=|\{j\in (1,\ldots, n ):  \theta_{ij}>\theta_{ij}^0\}|.$ If the alternative is left sided, then the inequalities above are reversed, and the number of false null hypotheses is defined as $n_i^-=|\{j\in (1,\ldots, n ):  \theta_{ij}<\theta_{ij}^0\}|$.
 If the alternative is two sided, then $H_{ij}: \theta_{ij}= \theta^0_{ij}$, and $H_{ij}$ is false if  $\theta_{ij}\neq \theta^0_{ij}.$ In this case, the number of false null hypotheses for feature $i$ is $n_i^++n_i^-.$ 

\begin{definition}[The $r$ out of $n$ replicability, $2\leq r\leq n$]\label{def:replic}
 If the alternative is right-sided, then we have $r/n$ replicability for feature $i$  if $n_i^+\geq r$.
The $r/n$ no replicability null hypothesis is   \begin{align}H_i^{r/n, +}: n_i^+\leq r-1, \label{part-conj-no-dir1}\end{align}
so we have $r/n$ replicability  if $H^{r/n, +}$ is false.  
If the alternative is two-sided, then  we have $r/n$ replicability for feature $i$ if $n_i^+\geq r$ or $n_i^-\geq r$, i.e., at least $r$ of the parameters in the $n$ studies addressing feature $i$ are non-null in the same direction. 
The $r/n$ no replicability null hypothesis is   \begin{align}H_i^{r/n}: \{n_i^+\leq r-1\} \cap  \{ n_i^-\leq r-1\}, \label{part-conj-dir1}\end{align}
 so we have $r/n$ replicability  if $H_i^{r/n}$ is false. 
\end{definition}

The \emph{minimal no replicability null hypothesis} for feature $i\in\{1, \ldots, m\}$ is given by (\ref{part-conj-no-dir1}) or (\ref{part-conj-dir1}) with $r=2$. The case $r=1$ reduces to the  global null hypothesis which is typically tested in  meta-analysis. However, rejection of the global null hypothesis does not provide evidence towards replicability since the rejection may be due to signal in a single study.

For $n=2,$ the only possible replicability claims are minimal replicability claims. However, for $n>2,$ one can address the question whether the $r/n$ replicability claim is true for different values of $r,$ specifically, $r\in\{2, \ldots, n\}.$ Obviously, the higher is the value of $r,$ the stronger is the corresponding replicability claim. For example, if three studies address the question whether a  certain relationship holds, a claim that this relationship holds in all the three studies is  stronger than a claim that it holds in at least two of the three studies. Thus, it may be attractive to test only the $n/n$ no replicability null hypotheses, rejection of which gives the strongest possible claims regarding replicability. However, the power to reject the $r/n$ no replicability null hypothesis decreases as $r$ increases (since the composite null is larger, $H_i^{r/n}\subset H_i^{(r+1)/n}$). Therefore, even if 
the $n/n$ no replicability null hypothesis is false, the power  may be too low to detect it in practice.   
It is worth to take these power considerations into account when making the choice of $r$ for $n>2.$ In \S~\ref{single} and \S~\ref{sec:frequentist}  we  show methods that do not require fixing $r$ in advance. These methods 
provide a lower bound on the number of studies with signal for each selected feature, with a confidence guarantee.

   Searching for replicated effects across studies has been considered in several application fields, where  different terms have been used for what we call $r/n$ replicability.  This is referred to as generalizability in \cite{Nosek22, Sofer17}, who consider generalization of effects to different ethnic groups. In the GWAS literature, it is referred to as pleiotropy when the phenotypes studied are distinct. For example, the single nucleotide polymorphisms (SNPs)  that are associated with more than one psychiatric disorder are pleiotropic SNPs (e.g., the SNPs associated with both Scizophrenia and bipolar disorder, \citealt{Andreassen13}). It is referred to as consistency or external validation when the different studies correspond to different environments \citep{li2021searching}.

Assessing replicability can be useful for increasing the evidence for causality of effects.
For example, higher rates of leukemia among radiologists and among survivors of Hiroshima and Negasaki provide replicable evidence of the increased risk of leukemia caused by radiation (\cite{Rosenbaum01}, see more examples in \cite{Rosenbaum22}).  For assessing causality in etiological epidemiology, \cite{lawlor16} have shown the usefulness of identifying that an effect is replicated in studies with different biases, some of which are, desirably, in different directions. \cite{lawlor16} use the term \emph{triangulation} to describe obtaining more reliable or accurate answers to research questions by comparing results in studies from two or more different epidemiological  approaches. Since triangulation will mostly provide a qualitative assessment of the strength of evidence regarding causality \citep{lawlor16},  the methods reviewed in this paper could be used in order to complement this assessment with quantitative measures regarding replicability strength across the different studies. 

Definition \ref{def:replic} is  useful   when methods of measurements differ and may not be on the same scale from one study to the next, since $\theta_{i1}^0$ need not equal $\theta_{i2}^0$. Moreover, it is useful when only the qualitative conclusion matters, so it is enough to know that the association/phenomenon exists in at least $r$ of the studies. \cite{Simonsohn15} argues that this is the case for many studies in psychology. As an example, he discusses the (hypothetical) phenomenon of levitation: if  one experiment concludes that people can levitate on average 9 inches above ground, and another concludes that the average is 0 inches, then the meta-analytic average is 4.5 inches but there is no replicated evidence for the phenomenon. There would have been 2/2 replicability only if  both experiments concluded that people can levitate on average a positive amount above ground (and this positive amount may differ across the studies). 

\begin{remark}
According to our definition, a true $r/n$ replicability claim is made if at least $r\geq 2$ of the parameters in the $n$ studies are non-null (in the same direction), even if the non-null parameters differ across studies. We realize that this definition of replicability may seem unsatisfying for some purposes. \cite{HedgesSchauer19b} argue that if effects are in the same direction but vary in several orders of magnitude, the result may not be considered replicated. If the primary study has a positive effect, and a follow-up study has a positive yet much smaller effect, they would like to define it as non-replicated. This is in contrast with our definition which, with enough data, will consider it replicated. One way to alleviate their concern, when the hypothesis testing problems are one sided,  is to define a feature as replicated if \emph{the effect is at least of a meaningful  (predefined) magnitude in  at least $r$ studies} \citep{Mathur19, jaljuli2022quantifying}. 
\end{remark}
 \begin{remark}
We concentrate on identification of the replicated signals. For assessing field-wise replicability measures we  refer the reader to \cite{Fithian-AoAS-2020} and references within. These measures are important when focusing on the reliability of a field of science. This is not the focus of this review, though it is related in the sense that by routinely applying the procedures discussed in this review, which aim to detect replicated signals while controlling for false replicability claims,  the reliability of the field may increase. Specifically,  the fraction of published  replicable relationships that are false is expected to be smaller  than the fraction of published relationships that are false. 
\end{remark}

\begin{remark}
One of the settings addressed in \S~\ref{sec:multiple studies} considers the   null hypothesis $H_{ij}$ to be  conditional independence between an outcome and a predictor $i$, given all other predictors, in study $j$ \citep{li2021searching}. With some abuse of notation, the $r$ out of $n$ replicability in this case is that $n^+_i\geq r$, where $n^+_i$ represents the number of studies in which predictor $i$ and the outcome are conditionally dependent. 
\end{remark}

\subsection{Reasons for lack of replicability}
Since there is unavoidable heterogeneity across studies, it can happen that the effect is present in exactly one of the studies, and therefore it is clearly not a replicated effect. Formally, $H_i^{r/n,+}$ may be true for $r\geq 2$, even though $n_i^+=1.$
This explains why rejecting the global null hypothesis, $n_i^+=0,$ 
is not enough in order to make a replicability claim. For $r/n$ replicability with $r>2$, even rejecting $H_i^{(r-1)/n,+}$ is not enough - it is necessary to reject $H_i^{r/n,+}$.

For example, in GWAS,  an association between genotype and phenotype may be particular to a single study and thus non-replicable. We provide the following documented case regarding Crohn's disease.  
 \cite{franke2010genome} has successfully detected more than 70  genomic regions which are susceptible for association with Crohn's disease in populations of European ancestry, but only a few have been confirmed    in the East Asian population \citep{Marigorta18, Nakagome12}.  As suggested by \cite{Nakagome12}, the lack of any association with Crohn's disease in a specific region NOD2 in East Asians could be due to the specificity of the causal association to the population of European ancestry due to natural selection. Another possible explanation is the lack of representation of causal variants, see Section 5.2 of \cite{li2021searching}. Briefly, the genome has block-like patterns of linkage disequilibrium, so that the alleles are dependent within each block, and are approximately independent across the blocks. For certain non-causal SNPs, the null hypothesis of conditional independence may be false just because they belong to the same block as the unmeasured causal SNP. The block-like patterns of linkage disequilibrium vary across populations with different ancestries \citep{Kidd04}. Therefore, a  non-causal SNP may be in the same block as the unmeasured causal SNP in the population addressed by  study one, implying the falsehood of its null hypothesis in study one, but in a  block containing no causal SNPs in a population addressed by study two, implying the truth of its null hypothesis in study two.

Additional examples of cases where a null hypothesis may be false in one study and true in the other study can be found in \cite{Wang20}, who search for replicated signals across different platforms or ethnic groups. Of course it may be that the signal is present only in a single study, which is carried out on a specific platform or on a specific ethnic group.  

  Another reason for a finding to fail to replicate is that the finding may be due to bias: the null hypothesis of no treatment effect is true, but its $p$-value is stochastically smaller than uniform, so it behaves as a $p$-value of a false null hypothesis. 
  This is certainly a danger in observational studies. If hidden bias is present,  then differing outcomes in treated and control groups may not actually be effects caused by the treatment. Suppose that this is the case in a single study, and that there is no treatment effect. Suppose also that the other  studies considered do not replicate the same hidden bias. Then  the prospect of replicating the finding (a `treatment effect') in the single study by the other studies is reduced \citep{Rosenbaum01}. This may be the case in  
  \cite{Hughes20, Panagiotou20}, where it is found  that  blood pressure lowering with antihypertensive agents  is associated with a lower risk of incident dementia or cognitive impairment in at least one study, but possibly only in one study.

\section{Error rates  for controlling erroneous replicability claims}\label{notation}

Let us first consider the case where the alternatives are   right sided. In general, our goal is to identify features with effects in at least $r$ studies, for $r\in\{2, \ldots, n\}.$ This goal can be achieved by testing the family $\{H_i^{r/n,+}, i=1, \ldots, m\}.$ Rejection of $H_i^{r/n,+}$ leads to claiming that an effect for feature $i$ exists in at least $r$ studies. This replicability claim is false if $H_i^{r/n,+}$ is true, i.e., if the number of studies with effects for feature $i$ is smaller than $r.$ Let   $V^{r/n}$ and $R^{r/n}$  be the number of false replicability claims and the total number of replicability  claims, respectively. 
The family-wise error rate (FWER) for replicability analysis is the probability that one or more replicability claims that were made are false, i.e. $$\text{FWER}^{r/n}=\mP(V^{r/n}>0).$$ Similarly, the false discovery rate (FDR, \cite{benjamini1995controlling}) for replicability claims is  $$\text{FDR}^{r/n}=\mE\left[\frac{V^{r/n}}{\max(R^{r/n}, 1)}\right].$$ Rather than identifying the features with effects in at least $r$ studies, one can obtain a lower confidence bound for the number of such features  (or equivalently, an upper confidence bound on the number of features that have an effect in  at most $r-1$ studies), in any selected set of features, with a pre-defined simultaneous coverage guarantee, see  \cite{goeman2011multiple} and \cite{blanchard2020post}, among others, for methods addressing this goal.

For two-sided alternatives it is of interest to identify features with effects in the same direction in at least $r$ studies, and to declare the common direction of effects.
Let $R_i^{r/n,L}$ and $R_i^{r/n,R}$ be the indicators of whether the $r/n$ replicability claim for feature $i$ was made in the left and right directions, respectively. Define $R^{r/n,D}=\sum_{i=1}^mR_i^{r/n,L}+\sum_{i=1}^mR_i^{r/n,R}$ as the total number of $r/n$ replicability claims. The directional replicability framework accounts for type III errors \citep{tukey1991philosophy, jones2000sensible},  which occur when the declared common direction of effects is wrong.  Specifically, the numbers of false and true directional $r/n$ replicability claims are $V^{r/n,D}$ and $S^{r/n,D},$ respectively, where
\begin{align*}&S^{r/n,D}=\\&\sum_{i: n_i^+\geq r}R_i^{r/n,R}+\sum_{i: n_i^-\geq r}R_i^{r/n,L} ,\end{align*} and $V^{r/n,D}=R^{r/n,D}-S^{r/n,D}.$ The FWER and FDR for directional $r/n$ replicability analysis are defined with respect to the above definitions of true and false discoveries, i.e.

$$\text{FWER}^{r/n,D}=\mP(V^{r/n,D}>0),$$
and $$\text{FDR}^{r/n,D}=\mE\left[\frac{V^{r/n,D}}{\max(R^{r/n,D}, 1)}\right].$$
The next section addresses the case where there is a single feature examined in each study ($m=1$), in which the error rates above reduce to the probability of making a false $r/n$  replicability claim, or  at least one false directional $r/n$ replicability claim. In \S~\ref{sec:frequentist} we address the case of multiple features, and review several methods for controlling the error rates above. In \S~\ref{sec:Bayesian} we address the control of  the Bayesian counterpart of the FDR on replicability claims, for high dimensional studies.

\section{The case of a single feature per study }\label{single}
\subsection{Parametric and naive approaches}
Systematic reviews attempt to assemble all the studies that are relevant to specific research questions.  They are important for advancing good standards of care  in  the field of medicine, where the questions are about  medical interventions. In particular,  the Cochrane collaboration routinely conducts systematic reviews in order to assess the effects of health care interventions \citep{Higgins22}. 

The systematic review may carry out one, or several, meta-analyses.  Each 
meta-analysis  carried out examines a single intervention. For a meta-analysis of a specific (e.g., primary) outcome in the systematic review, let 
$\boldsymbol{\theta_1} = (\theta_{11}, \ldots, \theta_{1n})$ denote the (unknown) effects of the intervention in the $n$ studies meta-analysed.
The meta-analysis   is used to infer on  whether the effect of the intervention  is present in at least one study (by rejecting the test of the global null hypothesis that there is no effect in all the studies). In addition, the meta-analysis  typically  provides confidence intervals assuming that the   effect  is common to all studies (the  `fixed effect' model, where it is assumed that $\theta_{11} = \ldots = \theta_{1n}$) or that the effects underlying the studies follow some distribution,
the `random effects' model. 

For the `random effects' model, the assumed underlying distribution is typically Gaussian,  so it is assumed that $\theta_{1j} \stackrel{iid}\sim N(\mu, \tau^2)$. With this strong assumption on the distribution of the  $\theta_{1j}$s,  it is possible to infer on the overall effect \citep{Higgins22}, as well as provide $1-\alpha$ level prediction intervals for the $\theta_{1j}$'s.  
\cite{Saad18}
view the intervention effect in a new study as independently sampled from the random-effects model. By projecting the estimates and confidence regions for the population mean and standard deviation, they produce point-wise estimates and confidence intervals for the cumulative distribution function of the random effects model, which provide a quantitative method for clinical decision-making that takes into consideration the heterogeneity of the intervention effect. Replicability can also be addressed within this model:  \cite{Mathur19} suggest several  metrics for inference, including the fraction of effects that exceed a predefined threshold.

The assumption that the true (unobserved) treatment effects come from a known distribution can have a detrimental impact on inference if it is false. This is so especially when the number of studies is small so the assumption cannot be verified using diagnostic tools. In particular, in systematic reviews of medical interventions, the number of studies is typically small, and  there is no reason to believe that the unobserved treatment effects in a few different cohorts (or subgroups, or environments) have a bell-shaped histogram that concurs with the model assumption.
Therefore, we would like to provide tools for assessing the extent of the evidence towards replicability, that do no rely on any parametric assumptions.

An intuitive way of  declaring $n/n$ replicability, which is also statistically valid, is if the $p$-values from all $n>1$ studies are at most $\alpha$ (the predefined probability of a type I error). For $n=2$, the probability of a false replicability claim is then at most $\alpha$, and this bound is tight since it occurs when the $p$-value in one study is almost surely zero (i.e., a very strong association is detected), yet in the other study the $p$-value is uniform. However, for $n>2$ the requirement that all $p$-values be at most $\alpha$ is too stringent, at least when it is enough to establish that there is signal in at least two  studies - the  minimal replicability requirement. 

A naive approach is to declare minimal replicability if at least two studies have $p$-values at most $\alpha$. However, the naive approach is not valid, since the probability that the replicability claim is false can be far greater than $\alpha$. To see this,  consider the common setting in which $p$-values corresponding to true null hypotheses are uniformly distributed, so the probability  that each of these $p$-values is at most $\alpha$ is $\alpha$. If at most one study is non-null,  then there is no replicability. However,  the probability that at least two studies will have $p$-value at most $\alpha$ is lower bounded by  the probability that a binomial random variable with $n$ trials and success probability $\alpha$ is at least two, i.e. $1-(1-\alpha)^{n}-n(1-\alpha)^{n-1}\alpha$. For example, for $n=10$ and $\alpha = 0.05$,  
it is 0.09. This lower bound is the probability of a false replicability claim (i.e., of rejecting $H^{2/n}_1$, even though $H^{2/n}_1$ is true),  only when all the $p$-values are uniformly distributed. However, if a single study has $p$-value less than $ 0.05$ with probability one, and the remaining $n-1=9$ $p$-values are uniformly distributed, then the probability of a false replicability claim increases to $1-(1-\alpha)^{n-1} = 1-0.95^9 = 0.37$. The naive approach should not be used for establishing replicability, since the probability of falsely declaring minimal replicability  is non-negligible even for $n$ small, and increases to one rapidly.

 In \S~\ref{subsec-PCpvalues} we briefly describe the approach in \cite{jaljuli2022quantifying}, which is  completely non-parametric, and therefore it  can be used for replicability analysis when $n$ is small or large.

\begin{remark}\label{rem-Benjamin}
Decreasing the rejection threshold from $\alpha=0.05$ to $\alpha=0.005$, following \cite{Benjamin18}, indeed decreases the probability of making a false replicability claim with the naive approach. However, when the number of studies $n$ is large enough, this probability can still be close to one. In \S~\ref{subsec-PCpvalues} we suggest a method that guarantees control of the probability of making a false replicability claim at the desired level $\alpha$. Of course, as $\alpha$ decreases the power to detect a false  no-replicability null hypothesis decreases, and the power deterioration may be large when moving from  an $\alpha=0.05$ to  $\alpha =  0.005$. The suggestion to reduce $\alpha$ by \cite{Benjamin18} was motivated by the need to have a stricter statistical standard for claiming new discoveries. Since the claim of replicability is stronger than the claim of a single new discovery, it may not be necessary to raise the standards even higher by  reducing $\alpha$.
\end{remark}

\subsection{Replicability analysis with partial conjunction $p$-values}\label{subsec-PCpvalues}

For a general method for obtaining a valid $p$-value for the $r/n$ no replicability null  hypothesis \eqref{part-conj-no-dir1}, see \cite{benjamini2008screening, Wang19}, where this null hypothesis and its $p$-value are referred to as $r/n$ \emph{partial conjunction} (PC) null hypothesis and $p$-value, respectively. 
One example for a PC $p$-value is the following. Let $p_{1j}$ be the right-sided $p$-value associated with $H_{1j},$  in study $j,$ for  $j=1,\ldots, n$. The $n$ $p$-values are mutually independent, since we address the setting of $n$ independent studies.  Let  $p_{1(1)}\leq p_{1(2)}\leq \ldots\leq  p_{1(n)}$ be the ordered sequence of right-sided $p$-values. The PC $p$-value motivated by Fisher's $p$-value combining method \citep{fisher1932statistical} for $H^{r/n,+}_1$ is 
\begin{align}p_1^{r/n}=\mP\left[\chi^2_{2(n-r+1)}\geq -2\sum_{j=r}^{n}\log p_{1(j)}\right]\label{fisher} \end{align}

\cite{jaljuli2022quantifying} use the PC $p$-values in order to establish replicability in systematic reviews from the Cochrane library. The $r/n$ PC $p$-value is a $p$-value for the $r/n$ no replicability null hypothesis, so the smaller it is the greater the evidence towards $r/n$ replicability. It is thus an informative quantitative measure of the extent of evidence towards replicability, just as the $p$-value is informative for reporting in individual studies. This measure is attractive for assessing replicability for the following reasons: 
\begin{itemize}
    \item Even if all studies are underpowered, it may convey that there is evidence towards replicability (this property is similar to the attractive property in meta-analysis of pooling evidence across the studies in order to discover that there is signal in at least one study). In particular, even if the $p$-values in all individual studies are larger than $\alpha = 0.05$, the PC $p$-value may be smaller than 0.05. For example,  for $n=3$ $p$-values with value 0.090, $p_1^{2/3} =0.047 $ with Fisher's combining method. 
    \item Even if $r-1$ studies have tiny $p$-values, it may convey that there is no evidence for $r/n$ replicability. For example, for $n=3$ $p$-values with values 0.00001, 0.2, and 0.4, $p_1^{2/3} =0.28 $ with Fisher's combining method; the global null $p$-value is $p_1^{1/n}  = 9\times 10^{-5}$. So in this example we can conclude that there is signal in at least one study, but  we cannot rule out that the signal may be present only in one study (since  we fail to establish minimal replicability at $\alpha=0.05$, or any level $\alpha<0.28$). 
\end{itemize}

In addition, a $1-\alpha$ confidence lower bound on the number of studies with effect is provided  by testing the PC hypotheses in order starting from $r=1,$ following \cite{BHY09}. Specifically, this lower bound is given by
\begin{align}l_\alpha=\max\{l\in \{0, \ldots, n\}: p_1^{r/n}\leq \alpha \text{  for  } r=0, \ldots, l\},\label{bound}\end{align} where $p_1^{0/n}$ is defined as 0 (see \cite{jaljuli2022quantifying} for details). 

For directional inference, one-sided $p$-values are combined in each direction in \cite{jaljuli2022quantifying}, following \cite{owen2009karl}. For example, the PC $p$-value for $ H_1^{r/n}$ using Fisher's combining method is $2\min \{q_1^{r/n}, p_1^{r/n} \}$, where $p_1^{r/n}$ is defined in \eqref{fisher} and  $$q_1^{r/n} = \mP\left[\chi^2_{2(n-r+1)}\geq -2\sum_{j=r}^{n}\log q_{1(j)}\right],$$ where $q_{1(1)}\leq q_{1(2)}\leq \ldots\leq  q_{1(n)}$ is the ordered sequence of left-sided $p$-values (for continuous test statistics, $q_{1j} = 1-p_{1j}, \ j=1, \ldots,n$). If the  PC $p$-value above is below $\alpha,$ the common direction of effects is declared to be right-sided if $p_1^{r/n}<q_1^{r/n},$ and left-sided otherwise.

\begin{remark}
 The methodology we describe takes care to control the probability of declaring replicability falsely in the case that exactly one study has an association, i.e.,  $n_1^+=1$ when testing right-sided alternatives.
 We believe it is important to apply replicability analysis methods that are valid  even when $n_1^+=1$,  since such methods guard against claiming replicability due to evidence coming only  from a single study, which is a realistic setting that cannot be ruled out, as explained in Section \ref{sec-definigreplicability}.  Of course, if it is a priori somehow known that  $n_i^+=1$ cannot be true, then it is possible to device more powerful criteria for replication success. The potential increase in power is due to the fact that in this case  the no replicability null hypothesis, $n_1^+=0$, is much narrower  than \eqref{part-conj-no-dir1}.  
 Specifically, for testing the 2/2 no replicability null hypothesis  we require that  $\max(p_{11},p_{12})\leq \alpha$ in order to declare replicability. The least favorable configuration, for which the probability that  $\max(p_{11},p_{12})\leq \alpha$  is exactly $\alpha$, is that $n_1^+=1$, and that this study has a $p$-value of zero (almost surely), and the other study has a $p$-value that is uniformly distributed. However, if it assumed that $n_1^+\in \{0,2 \}$,  then if there is no replicability, the global null is assumed to be true and so the probability that $\max(p_{11},p_{12})\leq \alpha$ is $\alpha^2$ (i.e., the test is too conservative). Under this strong 
 assumption, where no replicability means that the global null is true,  \cite{Pawel20, PawelHeld22, Held22} suggested less conservative tests for replication success.  
 \end{remark}

\subsection{Extensions and future directions}

The difference in power between different  combining methods used for testing $H^{r/n}_1$ can be large, and the combining methods differ in the settings in which they work best. A recent comparative study \citep{hoang2021combining} examines several combination methods of $p$-values as well as of $e$-values. Choosing a good combination method is important in order to have good power to establish replicability, but the choice depends on a-priori unknown characteristics of the data generation, notably  the number of studies with signal. Therefore, it can be useful to have adaptive methods that let the data guide the choice of the appropriate combination method for testing the no-replicability null hypothesis.

Meta-analysis is prone to publication bias, where only significant results (typically at $p$-value $\leq$ 0.05) are published. If the group of studies considered may be biased,  a sensitivity analysis that takes the bias into account should be carried out. For example, the $p$-values for the replicability analysis may be  adjusted for selection bias  by combining for PC testing only $p$-values at most 0.05, after inflating each by multiplying it by   $1/0.05=20$ \citep{fithian-arxiv, Zhao2019, Fithian-AoAS-2020}. Such an analysis is more conservative, but also more robust to the bias that arises from publishing only findings that are significant at the 0.05 level.   

A lower confidence bound on $n_1^+$ is of great interest for assessing replicability. An upper confidence bound may also be of  interest, since it  conveys the limit on replicability, or the specificity, to a subset of studies.  \cite{Heller2023} provides methods targeting upper (as well as lower) confidence bounds for $n_1^+$. 

Finally, the systematic reviews for health care interventions typically consider multiple primary and secondary outcomes. Therefore, they  carry out multiple meta-analyses that each examines a single endpoint. It can be thus useful to develop a replicability analysis pipeline that takes the multiplicity of the  outcomes into account, to guarantee a desired overall error rate. The setting of multiple outcomes fits into the methods we discuss next in its simplest form where the outcomes can be treated as exchangeable. This is not necessarily the case for medical interventions, where it  may be best to use weights or have a hierarchical structure that  conveys the relative importance of, or dependence between, the outcomes.

\section{Frequentist approaches for multiple features per study}\label{sec:frequentist}
\subsection{The shortcomings of   the naive approach }\label{subsec-invalid-naive}

A common practice is to  corroborate the reported findings from a study with reported  findings in one or more previous studies that examine the same or a similar problem. Although this is an intuitive approach for establishing replicability, it is  not  a formal statistical approach. For a single feature, i.e. $m=1,$ this approach reduces to the naive approach of declaring minimal replicability when the $p$-value is at most $\alpha$ in at least two studies among the available $n$ studies. As shown in Section \ref{single}, this approach is valid for $n=2,$ but it is not valid for $n>2.$ In this section we address the settings where $m>1,$ and characterize the scenarios in which the naive approach may be problematic in the following two aspects:  it may cause an unacceptable inflation of false replicability claims; it may lack  power to discover features with replicated signals  (that a non-naive replicability analysis procedure may have good power to discover).   

Let us first formalize the naive approach that we address. Although the investigators often report in their papers that their findings were replicated in one or two studies examining the same features,  in practice, it seems natural that the investigators consider multiple previous studies examining their features of interest, and pick for claiming replicability only those which indeed corroborate their findings. 
Our formal simplification is that the analyst declares as replicated the features in the set
$\mathcal{D}_1\cap (\cup_{j=2}^n \mathcal{D}_j),$ where $\mathcal{D}_1$ is the set of features discovered in the given study, call it study one, and $\mathcal{D}_j,\,j=2, \ldots, n$ are the sets of features discovered in $n-1$ previous studies, where each $\mathcal{D}_j$ is the result of a multiple testing procedure (e.g., Bonferroni or BH).
Of course in practice $n$ may  not be fixed in advance. 

We illustrate the problematic aspects  of the naive approach by examples below and in Appendix \ref{sec-SM}. Briefly, some key observations are as follows.  First, suppose that the proportion of non-replicable signals in study one  is non-negligible. Suppose that the discoveries in study one are those that pass the severe Bonferroni correction at level $\alpha$ (i.e., only features with $p_{i1} \leq \alpha/m$ are rejected). Then, for any fixed $m$ and $n>2$, the naive approach  may lead to an inflated FWER for minimal replicability analysis, i.e. the probability of making one or more false minimal replicability claims may be higher than $\alpha.$ The inflation may increase, and thus be of greater concern, as  (1) the number of studies considered, $n,$ increases, or (2) the proportion or strength of non-replicable signals in study one increases, or (3) the analyst applies a   multiple testing procedure  within each study that is less stringent than Bonferroni, e.g., if Bonferroni is replaced by  Hommel's procedure \citep{hommel1988stagewise} or the Benjamini-Hochberg (BH) procedure   \citep{benjamini1995controlling}. The naive approach with discoveries from the BH procedure may lead to a highly inflated FDR for replicability analysis even for $n=2$. Second, consider the case when  $m$ is large and the proportion of non-replicable signals in study one is close to 0, which is a typical scenario in GWA studies. Then the naive approach with Bonferroni and with BH may be valid in terms of FWER and FDR control, respectively. However, in these settings their power to discover features with replicated signals may be very low, compared to the  procedures in \S~\ref{sec: two studies} and \S~\ref{sec:multiple studies}, which are designed for controlling the FWER or FDR for replicability analysis. We thus recommend using the latter procedures and not the naive approach above, both for the sake of  replicability error control and for the sake of power for identifying true replicated signals. 

We now give more details and examples for  justifying the observations above. For simplicity, we address one-sided hypotheses. We start by two examples which apply the Bonferroni procedure in each study.  They serve to demonstrate the validity, yet low power, the naive approach can have for $n=2$ and for GWAS-like problems with $n>2$, as well as the inflation it can have in non GWAS-like problems. 
We provide in the SM additional examples that assess the naive approach when in each study discoveries are made using FWER-controlling procedures (including those which are less conservative than Bonferroni). 

\begin{example}\label{ex:FWER-n2}
Assume one faces $n=2$ studies examining $m>1$ features, and the discoveries within them are obtained using the level-$\alpha$ Bonferroni procedure, which  rejects a given hypothesis if its $p$-value is below $\alpha/m.$ In this case, the set of indices of features with replicability claims based on the naive approach is $\mathcal{D}_1\cap \mathcal{D}_2=\{i: \max(p_{i1}, p_{i2})\leq \alpha/m\},$ i.e. the rejection set of Bonferroni applied on $\{\max(p_{i1}, p_{i2}), i=1, \ldots, m\}. $
Since for any $i\in\{1, \ldots, m\},$  $\max(p_{i1}, p_{i2})$ is a valid $p$-value for $H_i^{2/2,+}$ defined in (\ref{part-conj-no-dir1}), and the Bonferroni procedure guarantees FWER control under any dependency among the $p$-values, in this case the naive approach guarantees FWER control with respect to replicability claims. Although this approach is valid, it may be much more conservative than the FWER-controlling Procedure \ref{Biometrika-FWER} with Bonferroni in Step 2, especially when the proportion of signals within each study is small, see Remark 2 in \cite{bogomolov2018assessing}.
\end{example}

\begin{example}\label{ex:FWER-big}
We now address the configuration in Table \ref{tab:FWER-big}. For simplicity, we address the idealized setting where the $p$-values for false null hypotheses are, almost surely, equal to 0. 
\begin{table}[h]
   \centering
   	\begin{tabular}{|*{6}{c|}}
			\hline
			& Study 1 &Study 2 &  Study 3 &$\ldots$ & Study $n$  \\\hline
			 1 & 0 & $U(0,1)$ & $U(0,1)$ & \ldots & $U(0,1)$ \\\hline
			
				2 & $U(0,1)$ & $U(0,1)$ & $U(0,1)$& $\ldots$ & $U(0,1)$ \\\hline
				$\ldots$ &$\ldots$&$\ldots$&$\ldots$&$\ldots$ &$\ldots$ \\\hline
				$m-1$ & $U(0,1)$ & $U(0,1)$ & $U(0,1)$ & $\ldots$ &$U(0,1)$\\\hline
					$m$ & 0 &  0 & 0 & $\ldots$ &  0\\\hline
	\end{tabular}
    \caption{Illustration of the setting addressed in Example \ref{ex:FWER-big}. The $ij$'th entry indicates the distribution of $p_{ij},$ for each $i\in\{1, \ldots, m\},$ $j\in\{1,\ldots, n\}:$ $0$ indicates that the given $p$-value is  0 with probability 1, $U(0,1)$ indicates that its distribution is uniform.}
    \label{tab:FWER-big}
\end{table}
 We assume  that the hypotheses within each study are tested using Bonferroni at level $\alpha.$ 
 In this case, an analyst can make  false minimal replicability claims for all the features except for feature $m$, for which the effects are replicated in all the studies. The probability of making  a false minimal replicability claim for feature 1 is $g_1(\alpha)=1-(1-\alpha/m)^{n-1},$ and the probability of making a false replicability claim for a given feature with index $i\in\{2, \ldots, m-1\}$ is $g_2(\alpha)=(\alpha/m)\left[1-(1-\alpha/m)^{n-1}\right].$  
Therefore, under independence within each study, the FWER for minimal replicability analysis is equal to
\begin{align*}&\mP(\text{at least one false minimal replicability claim})=\\&1-\left[1-g_1(\alpha)\right]\left[1-g_2(\alpha)\right]^{m-2}=\\&
1-(1-\alpha/m)^{n-1}\left\{1-(\alpha/m)\left[1-(1-\alpha/m)^{n-1}\right]\right\}^{m-2}. \end{align*}
When $m$ is large and $n<<m,$ so that $n/m$ is close to 0, the FWER for replicability analysis is close to 0. This conclusion suggests that the naive approach is valid for GWAS, since this setting can be viewed as a stylized GWAS example\footnote{Although in GWAS the number of SNPs examined is often higher than a million, it is commonly assumed that the effective number of independent SNPs is a million, which leads to choosing the threshold of $5\times 10^{-8},$ corresponding to the Bonferroni correction of the target FWER level $0.05,$ which accounts for the multiplicity of a million features.}. However, as will become clear in the following sections, there are more powerful approaches than the naive approach.  
For $\alpha=0.05$ and  $n=10,$ the FWER for replicability analysis is $5\times 10^{-7}$ for $m=10^6$, and it increases to  0.005 for  $m=100$. Moving away from the GWAS setting, for small $m$ and $n>>m,$ the FWER for replicability analysis may approach 1. For $\alpha=0.05,$ $m=10$ and $n=100,$ the FWER is 0.4. 
\end{example}

Let us now consider the case where the hypotheses within each study are tested using the level-$\alpha$ BH procedure.  A toy example in the SM shows that even in the low-dimensional case where $m=n=2,$ the FDR for replicability analysis may be $2\alpha.$ The following example shows that the replicability FDR of the naive approach  can be close to 1 in settings with large $m$ and a non-negligible proportion of features with non-replicable effects.  For simulations showing the inflation of replicability FDR of the naive approach with BH under more realistic settings with $n=2$ studies, see Figure 1 in \cite{bogomolov2018assessing} and Figure 3 in \cite{bogomolov2013discovering}. Essentially, these results show that when the proportion of features with effects in exactly one study is small, the FDR of the naive approach may be controlled, however its power may be very low compared to other methods designed for controlling FDR for replicability analysis (e.g. Procedure \ref{Biometrika-FDR} of \cite{bogomolov2018assessing}). When the proportion of non-replicable effects increases, the naive approach loses its control of FDR for replicability claims. Interestingly, these simulation results essentially concur with the results shown in Figure 1 of \cite{li2021searching} with respect to the naive approach in the model-X framework of \cite{candes2018panning}, which uses a knockoff-based method rather than BH within each study\footnote{In the simulation for Figure 1 of \cite{li2021searching}, the naive procedure declares $n/n$ replicability for features with discoveries obtained in all the $n$ studies. This approach does not control  FDR for $n/n$ replicability analysis when most conditional associations are inconsistent. It retains FDR control when most conditional associations are consistent, but in this case it is under powered compared to knockoff-based procedures tailored for replicability analysis with FDR control (described in Procedure \ref{Knockoffs} ).}.

\begin{example}\label{ex:large-FDR}
An investigator faces $n=2$ studies examining $m$ features, where $m$ is large.
Assume that within each study $j\in\{1, 2\},$ the following conditions hold: (1) the $p$-values are independent, (2) the $p$-values for features with an effect in study $j\in\{1,2\}$  have the same cumulative distribution function, denoted by $F_j,$ which satisfies the conditions of Theorem 1 of \cite{genovese2002operating} (3) the features with no effect have uniform $p$-values. Assume that replicability claims are made for features for which the null hypotheses were rejected by BH at level $\alpha$ in both studies. Let $m_{00},$ $m_{11},$ $m_{01}$ and $m_{10}$ denote the numbers of features with no effect in both studies, with an effect in both studies, with an effect only in study two, and with an effect only in study one, respectively. Let $\pi_{rs}=m_{rs}/m,$ $r,s\in\{0,1\}$ be the corresponding proportions among all the $m$ features. We consider the setting where   $\pi_{11}=0,$ while $\pi_{10}$ and $\pi_{01}$ are both positive.
According to Theorem 1 of \cite{genovese2002operating}, for large enough $m$,  applying the BH procedure at level $\alpha$  within each study $j\in\{1,2\}$ is approximately equivalent to rejecting the hypotheses with $p$-values less than a fixed threshold $u_j^*$, which is the solution to $F_j(u)=\beta_ju,$ with $\beta_1^*=[1-\alpha (\pi_{00}+\pi_{01})]/[\alpha (1-\pi_{00}-\pi_{01})],$ and  $\beta_2^*=[1-\alpha (\pi_{00}+\pi_{10})]/[\alpha (1-\pi_{00}-\pi_{10})].$
In this setting any replicability claim is false, and the numbers of false replicability claims for features with no effects in both studies, with an effect only in study one, and with an effect only in study two, are approximately  $Bin(m_{00}, u_1^*u_2^*),$ $Bin(m_{10}, F_1(u_1^*)u_2^*),$ and $Bin(m_{01}, u_1^*F_2(u_2^*))$ random variables, respectively. Therefore,  for replicability the FDR is equivalent to the FWER, and is approximately \begin{align*}&1-(1-u_1^*u_2^*)^{m_{00}}(1-F_1(u_1^*)u_2^*)^{m_{10}}(1-u_1^*F_2(u_2^*))^{m_{01}}. 
\end{align*} For very strong effects, corresponding to $p$-values which are equal to 0 almost surely, it holds $F_1(u_1^*)=F_2(u_2^*)=1,$ so $u_i^*=1/\beta_{i}^*,$ for $i=1,2.$ Obviously, the expression above may be close to 1. For example, for $\alpha=0.05,$ and $(\pi_{00}, \pi_{10}, \pi_{01})=(0.85,0.1, 0.05),$ we obtain $u_1^*
=0.0052,$ $u_2^*
=0.0026,$ so $\text{FDR}\approx 0.995$ for $m=10,000$ and $\text{FDR}\approx0.235$ for $m=500.$ 
For a more sparse setting, where $(\pi_{00}, \pi_{10}, \pi_{01})=(0.98, 0.01, 0.01),$ we obtain $u_1^*=u_2^*=0.00053,$ so $\text{FDR}\approx 0.102$ for $m=10,000$ and $\text{FDR}\approx 0.005$ for $m=500.$ 
\end{example}

We review below several methods tailored for controlling FWER or FDR for replicability analysis, which we recommend using instead of the naive approach discussed above.
\subsection{Targeting minimal replicability}\label{sec: two studies}
 The goal of replicability analysis across two independent studies with summary statistics available for all the features has been considered in \cite{bogomolov2018assessing,zhao2018cross}. Their methods are based on two steps. First, the promising features are selected from each study separately, based on the data from that study. Second, the features that are selected from both studies are tested for replicability.  The methods are based on the idea that replicable features will seem to be promising in both studies if there is enough power to detect them, so (almost) all replicable features are expected to be selected for testing, while accounting for less features at the testing stage leads to power gain. Indeed, the method targeting FDR control is shown to be typically more powerful than  applying the BH procedure on maximum of the two studies' $p$-values (in case $n=2,$ the latter is the partial conjunction approach of \cite{BHY09}).
    
       For replicability with $\text{FWER}^{2/2, D}$ control at level $\alpha,$ the method  is the following. 
      \newlist{steps}{enumerate}{1}
\setlist[steps, 1]{wide=0pt, leftmargin=\parindent, label=Step \arabic*:, font=\bfseries}
\begin{algo}[\cite{bogomolov2018assessing, zhao2018cross}]\label{Biometrika-FWER} $ $
\begin{steps}
       \item[Step 1:] Apply a selection rule on the set of $p$-values $\{(p_{i1}, q_{i1}), i=1, \ldots, m\}$ to obtain the set of selected features $\mathcal{S}_1.$ Similarly, apply a selection rule on the set of $p$-values $\{(p_{i2}, q_{i2}), i=1, \ldots, m\}$ to obtain the set of selected features $\mathcal{S}_2.$  If all the alternatives are two-sided, define 
          \begin{equation*}
  p'_{i1} =
  \begin{cases}
    q_{i1}, & \,\,q_{i2}<p_{i2} \\
    p_{i1}, & \,\,q_{i2}>p_{i2} \\
  \end{cases},
  \,\,\,\,p'_{i2} =
  \begin{cases}
    q_{i2}, & \,\,q_{i1}<p_{i1} \\
    p_{i2}, & \,\,q_{i1}>p_{i1} \\
  \end{cases}
  \end{equation*}
       If all the alternatives are right-sided, define 
      $p'_{i1} = p_{i1},$ and $p'_{i2} = p_{i2}.$ 
       \item[Step 2:] Apply a FWER-controlling procedure (e.g. Holm's method, \cite{holm1979simple}) on the set $\{p'_{i1}, i\in \mathcal{S}_2\}$ at level $\alpha/2,$ and on the set
        $\{p'_{i2}, i\in \mathcal{S}_1\}$ at level $\alpha/2.$
          Let $\mathcal{R}_j$ be the set of indices of rejected hypotheses in set $\mathcal{S}_j$.
        \item[Step 3:]  Declare as replicated the features with indices in the set $\mathcal{R}_1\cap \mathcal{R}_2.$ 
      \end{steps}
      \end{algo}

      For replicability with $\text{FDR}^{2/2, D}$ control at level $\alpha,$
      \cite{bogomolov2018assessing} suggested the following method.
       Throughout the paper, for any event $A,$ $I(A)$ denotes the indicator of $A$, and inequality $x\leq y$ for vectors $x,y\in \mathbb{R}^n$ means that the corresponding inequality holds coordinate-wise, i.e. $x_i\leq y_i$ for $i=1, \ldots,n.$ 
      \begin{algo}[\cite{bogomolov2018assessing}]\label{Biometrika-FDR} $ $
      \begin{steps}
      \item[Step 1:] Identical to Step 1 in Procedure \ref{Biometrika-FWER}.  
       \item[Step 2:] Compute
       \begin{align*}&R=\\&\max\left\{r: \sum_{i\in \mathcal{S}_1\cap \mathcal{S}_2} I\left[(p'_{i1}, p'_{i2})\leq \left(\frac{r\alpha}{2|\mathcal{S}_2|}, \frac{r\alpha}{2|\mathcal{S}_1|}\right)\right]=r\right\} \end{align*}
        \item[Step 3:] Declare minimal replicability for features with indices in the set $$\mathcal{R}=\left\{i:(p'_{i1}, p'_{i2})\leq \left(\frac{R\alpha}{2|\mathcal{S}_2|}, \frac{R\alpha}{2|\mathcal{S}_1|}\right), i\in \mathcal{S}_1\cap \mathcal{S}_2\right\}.$$ 
      \end{steps}
      \end{algo}

  Rather than fixing the target replicability error rate $\alpha$ in advance, one can compute the  adjusted $p$-value for each feature, which quantifies its evidence for replicability. 
  The adjusted $p$-value for each feature is the minimal level of FWER or FDR for which the feature is declared as replicated by a given procedure.  Therefore, for error control at level $\alpha$,   
  replicability is established for features with adjusted $p$-values at most $\alpha.$

The adjusted $p$-values for the FWER-controlling Procedure \ref{Biometrika-FWER} depend on the choice of the FWER-controlling procedure used in Step 2. If Bonferroni's method is used in Step 2, then the adjusted $p$-value for feature $i\in\mathcal{S}_1\cap \mathcal{S}_2$ based on Procedure \ref{Biometrika-FWER} is
  \begin{align}\label{Bon-Biometrika}
      p_i^{\text{Bon}}=2\max(|\mathcal{S}_2|p'_{i1}, |\mathcal{S}_1|p'_{i2}).
  \end{align}
  The adjusted $p$-value for feature $i\in\mathcal{S}_1\cap \mathcal{S}_2$ based on Procedure \ref{Biometrika-FDR} is
  \begin{align}\label{FDR-Biometrika}
      p_i^{\text{BH}}=\min_{\{j:p_j^{\text{Bon}}\geq p_i^{\text{Bon}}, \,\,\,j\in \mathcal{S}_1\cap \mathcal{S}_2\}}\frac{p_j^{\text{Bon}}}{\sum_{k\in \mathcal{S}_1\cap \mathcal{S}_2}I(p_k^{\text{Bon}}\leq p_j^{\text{Bon}})},
  \end{align}
  where $p_i^{\text{Bon}}$ for $i\in \mathcal{S}_1\cap \mathcal{S}_2$ is given in (\ref{Bon-Biometrika}). 
      
         See \cite{bogomolov2018assessing} for theoretical guarantees of these procedures.   A straightforward generalization of these procedures for the case where $n\geq 2$ and the goal is identifying features with minimal replicability is as follows. Prior to Step 1, the $n$ studies are divided into two groups, and  the right-sided and left-sided $p$-values within each group are combined for obtaining  global null $p$-values for both directions. Then the same steps are applied, where $(p_{i1}, q_{i1})$ and $ (p_{i2}, q_{i2})$ are replaced by  the corresponding  global null  $p$-values for each group of studies. For both procedures, Step 3 defines the set of features with minimal replicability claims. The theoretical guarantees for the generalized Procedure \ref{Biometrika-FWER} and \ref{Biometrika-FDR} address $\text{FWER}^{2/n, D}$ and $\text{FDR}^{2/n, D},$ respectively. Interestingly, these procedures provide a partial identification of studies with effects for features with minimal replicability claims: for each such feature the global null in each of the two groups of studies is rejected, 
         i.e. one can claim that the feature has an effect in at least one study in each group. The lists of global null discoveries for each group come with an  $\alpha/2$-FWER control guarantee for generalized Procedure \ref{Biometrika-FWER}, and with an $\alpha/2$-FDR control guarantee for generalized Procedure \ref{Biometrika-FDR}.

         Variants of Procedures \ref{Biometrika-FWER} and \ref{Biometrika-FDR}, obtained by incorporating plug-in estimators for the proportions of nulls in one study among the selected in the other study, were suggested by \cite{bogomolov2018assessing} for gaining power. 
 Similar methods for the case where  one study  is primary and the other  one is a follow-up study, have been suggested in \cite{bogomolov2013discovering}, \cite{heller2014deciding}.
 
 \subsection{General replicability analysis}\label{sec:multiple studies}
  In case where there are more than two studies, one can address the goal of identifying features with effects in at least $r$ out of $n$ studies, for $r\geq 2.$ For $r=2,$ this reduces to minimal replicability, and the higher  $r$ is,  the stronger is the requirement for claiming replicability. 
  \cite{benjamini2008screening, BHY09}  suggested addressing this goal by multiple testing of $r/n$ no replicability null hypotheses. 
  Specifically, for $\text{FDR}^{r/n}$ control, they suggested applying the 
  BH procedure on $r/n$ PC $p$-values (see (\ref{fisher}) for an example of a PC $p$-value).  
  In some applications it may be unclear what is the appropriate choice of $r$ in order to claim replicability. For example, while one investigator may claim that association of  a genotype with a disease in at least two out of six studies is  enough for claiming replicability, another investigator may claim that stronger evidence is required, e.g. association with a disease in at least half of the studies (i.e. fixing $r=3$ rather than $r=2$). In order to provide the highest (non-zero) lower bound on the number of studies with effect for each feature, while controlling for non-coverage,  
  \cite{BHY09} suggested the following procedure for right-sided alternatives.
  \begin{algo}[\cite{BHY09}]\label{BHY09}$ $
  \begin{steps}
       \item[Step 1:] For each feature $i=1, \ldots, m,$ compute the global null $p$-value $p_i^{1/n}.$
        \item[Step 2:] Apply the BH procedure at level $\alpha$ on $\{p_i^{1/n}, i=1, \ldots, m\}.$ Select the features for which rejections are made. Let $\mathcal{S}$ be the set of indices of selected features.
        \item[Step 3:] For each feature $i\in \mathcal{S},$ test the no replicability null  hypotheses in order at level $|\mathcal{S}|\alpha/m,$ and define 
  $$ l_i=\max\{l\in \{0, \ldots, n\}: p_i^{r/n}\leq |\mathcal{S}|\alpha/m \text{  for  } r=0, \ldots, l\}$$
  to be the lower bound for the number of studies where feature  $i$ has an effect. 
      \end{steps}
      \end{algo}
 This procedure is a generalization of the method in \eqref{bound} for $m=1.$ \cite{BHY09}  proved that under independence, Procedure \ref{BHY09} guarantees that the expected proportion of features with lower bounds exceeding the true number of studies where they have effects, out of all the selected features, is upper bounded by $\alpha.$ 
  We refer to the error measure above as the false coverage rate (FCR, \cite{BY05}) for the selected features. 
  This procedure was generalized in \cite{Bogomolov21}, allowing for different rules for selecting the features for which lower bounds will be obtained. For example, the selected set $\mathcal{S}$ can be obtained by applying a multiple testing procedure on $p$-values of a given study, defining the discoveries in that study. Suppose a researcher applies Bonferroni for obtaining discoveries in his or her study, referred to as study one, then the selected set of features of interest regarding their replicability extent can be  $\mathcal{S}=\{i: p_{i1}\leq \alpha/m\}.$  The same can be done if the researcher's study is primary and the other studies are follow-up studies, examining only the features with significant  effects in the primary study. The generalized procedure is the following:
  \begin{algo}[\cite{Bogomolov21}]\label{proc-Bogomolov21}$  $
  \begin{steps}
  \item Apply a certain selection rule on all the $p$-values, for obtaining the selected set of features $\mathcal{S}.$
  \item Apply Step 3 of Procedure \ref{BHY09} with respect to the selected set of features $\mathcal{S}.$
      \end{steps}
  \end{algo}
It was shown in \cite{Bogomolov21} that the generalized procedure controls the FCR for the selected features  
under certain forms of positive dependencies within  the studies, and under lenient conditions on the selection rule used in Step 1 of Procedure \ref{proc-Bogomolov21}. In addition, the procedure can be conservatively adjusted for allowing arbitrary dependencies within the studies and arbitrary selection rules.
 
 A drawback of methods which are based on applying a multiple testing procedure on PC $p$-values is their possibly low power. The power loss occurs due to the fact that  $H_i^{r/n}$ is a composite null hypothesis. So when $H_i^{r/n}$ is true, the PC $p$-value gives an exact test only under the least favorable configuration, where the number of false null hypotheses for feature $i$ is $r-1,$ their corresponding $p$-values are all equal to 0, and the remaining $n-r+1$ $p$-values have a uniform distribution. Under other configurations, the PC $p$-value gives a conservative test. 
 For example, a very conservative, yet typically prevalent, configuration among the features is that the global null is true. 
 For feature $i$ with $n_i^+=n_i^-=0,$ i.e. for which the global null is true, $\mP(p_i^{n/n}\leq x)\leq x^n$.

 In order to overcome the power loss induced by the conservativeness of the PC $p$-values, \cite{Wang20} developed the AdaFilter methods for testing the family of PC hypotheses $H_1^{r/n,+}, \ldots, H_m^{r/n,+}$ for any given $r\in\{2, \ldots, n\}.$ Essentially, rather than adjusting for the multiplicity of all the $m$ features, these methods adjust only for the number of features with evidence for effects in at least $r-1$ studies, thereby gaining power. Specifics follow.
 Let $p_{i(1)}\leq p_{i(2)}\ldots \leq p_{i(n)}$ be the sorted right-sided $p$-values for feature $i\in\{1, \ldots, m\}.$ For testing the family  $H_1^{r/n,+}, \ldots, H_m^{r/n,+},$ the method is based on assigning the filtering statistic $F_i=(n-r+1)p_{i(r-1)}$ and the selection statistic $S_i=(n-r+1)p_{i(r)}$
 for feature $i\in\{1, \ldots, m\}.$
 The AdaFilter Bonferroni and AdaFilter BH, which are a version of AdaFilter targeting FWER control and FDR control, respectively, for the family $H_1^{r/n,+}, \ldots, H_m^{r/n,+},$ are
as follows.
 \begin{algo}[\cite{Wang20}, Adafilter-Bonferroni]\label{algo-Adafilter-FWER} $ $
 \begin{steps}
 \item Compute $$\gamma_0^{Bon}=\sup\Big\{\gamma\in[0, \alpha]\,\Big|\,\gamma\sum_{i=1}^m I(F_i< \gamma)\leq \alpha\Big\}.$$
 \item Reject $H_i^{r/n}$ if $S_i<\gamma_0^{Bon},$ for $i\in\{1, \ldots, m\}.$
 \end{steps}
 \end{algo}

  \begin{algo}[\cite{Wang20}, Adafilter-BH]\label{algo-Adafilter-FDR} $ $
 \begin{steps}
 \item Compute $$\gamma_0^{BH}=\sup\Big\{\gamma\in[0, \alpha]\,\Big|\,\frac{\gamma\sum_{i=1}^m I(F_i<\gamma)}{\max(\sum_{i=1}^mI(S_i<\gamma), 1)}\leq \alpha\Big\}.$$
 \item Reject $H_i^{r/n}$ if $S_i<\gamma_0^{BH},$ for $i\in\{1, \ldots, m\}.$
 \end{steps}
 \end{algo}

 For theoretical guarantees of the procedures, see \cite{Wang20}. 
 The AdaFilter Bonferroni adjusted $p$-value for $H_i^{r/n}$ is \begin{align}\label{Adafilter-Bon-adjusted}p_{i}^{Bon}=S_{i}\sum_{k=1}^m I(F_k\leq S_i)\end{align} and AdaFilter BH adjusted $p$-value for  $H_i^{r/n}$ is $$p_{i}^{\text{BH}}=\min_{
 \{j: p_j^{\text{Bon}}\geq p_i^{\text{Bon}}\}}\frac{p_j^{\text{Bon}}}{\sum_{k=1}^m I(p_k^{\text{Bon}}\leq p_j^{\text{Bon}})},$$
 where $\{p_i^{\text{\text{Bon}}}, i=1, \ldots, m\}$ are given in  (\ref{Adafilter-Bon-adjusted}).
 For any level $\alpha\in(0,1],$ the sets of rejections of AdaFilter Bonferroni and AdaFilter BH are $\{i: p_i^{\text{Bon}}<\alpha\}$ and $\{i: p_i^{\text{BH}}<\alpha\},$ respectively, as shown in Proposition 3.4 of \cite{Wang20}.

 \begin{remark}\label{rem:directional}
 For two-sided hypotheses and directional $r/n$ replicability for a certain fixed $r\in\{2, \ldots,n\},$ \cite{Wang20} suggested using their methods as follows. Apply the given method with parameter $r$ twice, separately on $\{p_{ij},i=1, \ldots,m,\, j=1, \ldots, n\}$ and on $\{q_{ij},i=1, \ldots,m,\, j=1, \ldots, n\},$ each time at level $\alpha/2.$ Let $\mathcal{R}^R$ and $\mathcal{R}^L$ be the corresponding sets of discoveries. Claim $r/n$ replicability of positive effect for features with indices in $\mathcal{R}^R,$ and $r/n$ replicability of negative effect for features with indices in $\mathcal{R}^L.$ This strategy guarantees control of directional FWER/FDR for replicability analysis as long as the corresponding method applied on one-sided $p$-values guarantees FWER/FDR for non-directional replicability. The approach of \cite{BHY09}, which amounts to applying BH on $r/n$ PC $p$-values, can be used similarly for directional $r/n$ replicability analysis.
 \end{remark}
 
 All the methods above address the general case where one has a $p$-value for each null hypothesis $H_{ij},$ which requires knowing the distribution of test statistics under the null.   \cite{li2021searching} addressed the goal of replicability in a different setting, which does not require  computing $p$-values. In their setting  one has observations  $(X, Y)$ consisting of $m$ variables $X\in \mathcal{X}^m$ and an outcome $Y\in \mathcal{Y}$ sampled from $n$ environments. 
 For variable $i$ and environment $j,$  the null hypothesis is defined as follows:
 $$H_{ij}: Y^j\ind X_i^j\,|\, X_{-i}^j,$$
 where $Y^j, X_i^j$ and $X_{-i}^j$, respectively, denote the outcome, the $i$th variable, and the vector of all variables except $X_i$ in environment $j.$ In words, $H_{ij}$ states that the outcome is independent of the $i$th explanatory variable conditional on all other explanatory variables, when considering the distribution of the data in the $j$th environment.  The alternative is complementary to the null, so it has no one-sided and two-sided variants. Therefore, directional replicability is not defined in this case.  
 Testing the $r/n$ no replicability null hypotheses for all $m$ variables addresses the goal of identifying conditional associations which hold in at least $r$ environments. The method of \cite{li2021searching} controls FDR for the family of $r/n$ no replicability null hypotheses. It relies on the model-X approach of \cite{candes2018panning}, which does not require the model for conditional distribution of the outcome, but requires that the distribution of $X$  within environment $j$ is known, at least approximately. The latter assumption is realistic in genetic studies, because reliable knowledge is available regarding the distribution of the genotypes \citep{sesia2019gene}. 
 The method relies on multi-environment knockoff statistics defined below, which generalize the knockoff statistics of 
 \cite{candes2018panning}. 
 \begin{definition}[\cite{li2021searching}]
 We say that $W\in \mathbb{R}^{m\times n}$ are multi-environment knockoff statistics   if $W$ has the same distribution as $W\cdot\epsilon,$ where $\cdot$ indicates element-wise multiplication, and $\epsilon$ is a random $m\times n$ matrix with independent entries in the set $\{1, -1\},$ satisfying $\epsilon_{ij}=\pm1$ with probability 1/2 if $H_{ij}$ is true, and $\epsilon_{ij}=1$ otherwise, for $i\in\{1, \ldots, m\},$ $j\in\{1, \ldots, n\}.$
 \end{definition}
 See \cite{li2021searching} for methods which can be used for constructing the multi-environment knockoff statistics. Given a fixed $r\in\{1, \ldots, n\},$ the method of \cite{li2021searching} for controlling the FDR for a family of $r/n$ no replicability null hypotheses is given below.
 \begin{algo}[\cite{li2021searching}]\label{Knockoffs}$ $
 \begin{steps}
 \item Compute the multi-environment knockoff statistics $W$. For each $i\in\{1, \ldots, m\},$ let $t_i^-=\sum_{j=1}^n I(W_{ij}<0),$ and $t_i^0=\sum_{j=1}^n I(W_{ij}=0).$ 
 \item For each $i\in\{1, \ldots, m\},$ compute \begin{align}p_i^{r/n}=\mP(T_i\leq t_i^--1)+U_i\,\mP(T_i=t_i^-),\label{Li-p-r}\end{align} where $T_i\sim  Bin(\max\{n-r+1-t_i^0,0
 \}, 1/2),$ and $U_i$ is a $U(0,1)$ random variable which is independent of everything else.
 \item For each $i\in\{1, \ldots, m\},$ consider the set of absolute values of knockoff statistics for variable $i$ in all the environments, $\{|W_{ij}|, j=1, \ldots, n\}.$ Obtain $|W_i^r|,$ defined as the product of the $r$ largest values in this set.  See \cite{li2021searching} for a more general definition of $|W_i^r|.$
 \item Filter the $p$-values obtained in Step 2 in decreasing order of $|W_i^r|$ using the level-$\alpha$ selective seqStep+ procedure of \cite{barber2015controlling} with a certain choice of the tuning parameter $c$, and obtain the rejected $r/n$ no replicability null hypotheses. Specifically, compute 
 \begin{align*}&\hat{w}=\\&\min\left\{w: \frac{1+|\{i: |W_i^r|\geq w, p_i^{r/n}>c \}|}{\max(|\{i: |W_i^r|\geq w, p_i^{r/n}\leq c \}|, 1)}\leq \frac{1-c}{c}\alpha\right\},\end{align*} and reject the $r/n$
hypotheses for variables with indices in the set $\{i: |W_i^r|\geq \hat{w}, p_i^{r/n}\leq c \}.$ 
 \end{steps}
  \end{algo}
See \cite{li2021searching} for another variant of the above procedure and its theoretical guarantees. A simpler method is suggested for the case where $r=n,$ i.e. when the goal is to find variables which are conditionally associated with the outcome in all the environments. Specifically, in Procedure \ref{Knockoffs} the definition in (\ref{Li-p-r}) for the PC $p$-value in Step 2 is replaced by the following definition: $p_i^{n/n}=1/2$ if $\min\{\text{sign}(W_{ij})\}_{j=1}^n=+1,$ and $p_i^{n/n}=1$ otherwise. The next steps are obtained by substituting $r=n$ in Steps 3 and 4, and choosing $c=1/2$ in Step 4.   
\subsection{Broad guidelines and future directions}
The properties of the replicability analysis methods addressed in this section are summarized in Table \ref{table-summary}. Let us address the comparison of Procedures \ref{Biometrika-FWER}, \ref{Biometrika-FDR} with Procedures \ref{algo-Adafilter-FWER}, \ref{algo-Adafilter-FDR} for the case where the goal is minimal replicability, i.e. $r=2.$ All these procedures test for replicability only the features that are selected as promising for assessing minimal replicability, however the selection mechanisms are different. While Procedures \ref{Biometrika-FWER}, \ref{Biometrika-FDR} allow the researcher to choose the selection rules within each study, Procedures \ref{algo-Adafilter-FWER}, \ref{algo-Adafilter-FDR} are based on fixed selection rules. In addition, Procedure \ref{Biometrika-FWER} allows using any FWER-controlling procedure in Step 2, on selected one-sided $p$-values, while Procedure \ref{algo-Adafilter-FWER} does not have this flexibility. The generalized versions of Procedures \ref{Biometrika-FWER} and \ref{Biometrika-FDR} for the case $n>2$ have an additional advantage over Procedures \ref{algo-Adafilter-FWER}, \ref{algo-Adafilter-FDR}: the former procedures allow partial identification of studies where the discovered features have effects, with an appropriate error rate control.  The disadvantage of generalized Procedures \ref{Biometrika-FWER}, \ref{Biometrika-FDR} is that their results depend on the division of the $n$ studies into two groups, which can be arbitrary. These procedures aim to identify features with effects in both groups of studies. 
Thus, for the case where $n>2,$ generalized Procedures  \ref{Biometrika-FWER}, \ref{Biometrika-FDR} may be less powerful than Procedures \ref{algo-Adafilter-FWER} and \ref{algo-Adafilter-FDR}, respectively, for the case where many features have two or more effects in only one group of studies. However, it may be possible to overcome this limitation by considering adaptively chosen groups. 
Moreover, it may be of great interest to further generalize Procedures \ref{Biometrika-FWER}, \ref{Biometrika-FDR}   for assessing $r/n$ replicability across $n$ studies for any $r\geq 2,$ thus carrying over to $r>2$ the flexibility and potential power gain of these procedures observed for $r=2$.

Procedures \ref{BHY09}, \ref{proc-Bogomolov21} are different from the other procedures, because rather than requiring to fix $r$ in advance, these procedures use the data for identifying what is the strongest replicability claim that can be made for a given feature. The replicability claims are made in terms of lower bounds for the number of studies with effects for the selected features, and the error rate controlled is FCR for the selected features.  As shown in Table \ref{table-summary}, the other methods  control for erroneous $r/n$ replicability claims for a fixed value of $r.$ If the adaptivity of Procedures \ref{BHY09}, \ref{proc-Bogomolov21} is attractive, then these procedures should be chosen for replicability analysis.

Procedures \ref{BHY09}, \ref{proc-Bogomolov21}, as well as Procedures \ref{algo-Adafilter-FWER}, \ref{algo-Adafilter-FDR} are given above for the case of one-sided hypotheses. Remark \ref{rem:directional} suggests a way to perform directional $r/n$ replicability using Procedures \ref{algo-Adafilter-FWER}, \ref{algo-Adafilter-FDR}, or using the partial conjunction approach of \cite{BHY09},  which amounts to applying the BH procedure on $r/n$ PC $p$-values. 
The latter  procedure is equivalent to that in Definition 7 of \cite{Benjamini05} for $n=1$. \cite{Benjamini05} show that the procedure that applies the BH procedure to the two-sided $p$-values and further declares the sign (their Definition 6) also controls the directional FDR, and is at least as powerful as their procedure in Definition 7. This suggests that the directional inference in Remark \ref{rem:directional} may be improved for directional FDR control on false replicability claims.  
An additional potential improvement may address overcoming the conservatism of the PC $p$-values used in these procedures.

Procedure \ref{Knockoffs} addresses a specific setting where one is given data for  $m$ explanatory variables and a certain outcome in each of $n$ environments (studies), and for each explanatory variable one is interested in testing the null hypothesis of  conditional independence between this variable and the outcome, given the other variables. This setting does not require specifying the model of dependence between the outcome and the explanatory variables.   
In contrast to other procedures in Table \ref{table-summary}, Procedure \ref{Knockoffs} is based on knockoff statistics and does not require the ability to compute $p$-values for the tested hypotheses.

All the procedures assume that the studies are based on independent data. This assumption does not always hold: it is obviously violated for case-control studies if  each study has its own disease cohort, but all are compared to the same control cohort; or if each study examines a different phenotype, but the same subject may be in several studies.  In these settings, it is therefore necessary to design novel powerful replicability analysis   procedures.

\begin{table}[!h]
\begin{center}
\begin{tabular}{p{0.05\textwidth} p{0.13\textwidth}p{2.3cm}p{1.5cm} }
   \hline
\centering{ Proc.}  & $r$ & Target error rate & Summary statistics \\
 \hline\ref{Biometrika-FWER} &$r=2$&   FWER$^{2/n, D}$ & $p$-values  \\\ref{Biometrika-FDR}  & $r=2$& FDR$^{2/n, D}$&   \\
 \hline
\ref{BHY09}, \ref{proc-Bogomolov21} & $r$ is not fixed in advance, adaptive & FCR on the number of studies with signal &  $p$-values  \\\hline
 \ref{algo-Adafilter-FWER}    & $r\geq 2$& FWER$^{r/n}$&  $p$-values  \\
 \ref{algo-Adafilter-FDR}   &$r\geq 2$ & FDR$^{r/n}$ &  \\\hline
 \ref{Knockoffs} & $r\geq 2$  & FDR$^{r/n}$ & knockoff statistics \\
 \hline
\end{tabular}
\caption{Summary for the replicability analysis procedures addressed in this section. The table addresses the generalized versions of Procedures  \ref{Biometrika-FWER}, \ref{Biometrika-FDR}, which can analyze $n\geq 2$ studies. The first, second, and third column give the reference to the procedure, the possible values of $r$ for which $r/n$ replicability can be assessed, and the target error rate addressing erroneous replicability claims, respectively. The fourth column gives the summary statistics required as input for each feature, within each study.}\label{table-summary}
\end{center}
\end{table}

\section{The empirical Bayes approach for replicability in high dimensional studies}\label{sec:Bayesian}

\subsection{Replicability analysis with the local false discovery rate}\label{subsec-eBayes}
The two group model \citep{Efron01, Efron10} is popular for analyzing high dimensional studies, where many features are simultaneously tested for association with an outcome. It has been generalized for targeting the discovery of features that are associated with the outcome in at least two studies in \cite{Chung14, heller2014replicability}.

The generalization of the two group model for $n$ studies, each examining a large number $m$ of features, is as follows. Using GWAS as our illustrative example, the indicator of the hypothesis state for feature $i\in \{1,\ldots,m \}$ in study $j\in \{1,\ldots,n\}$, $h_{ij}$,  is zero if there is no association, -1 if the association is negative, and one if the association is positive. The $z$-score $z_{ij} = \Phi^{-1}(p_{ij})$ has density $f_{j,1}(\cdot)$ if $h_{ij}=1$, $f_{j,0}(\cdot)$ if $h_{ij}=0$, and $f_{j,-1}(\cdot)$ if $h_{ij}=-1$. The null density $f_{j,0}(\cdot)$ is usually assumed to be the standard normal distribution (following the assumption that  $p_{ij}$ has a uniform null distribution). Each feature has probability  $\pi({\boldsymbol h})$ of having the vector of hypotheses states ${\boldsymbol h} \in \{-1,0,1 \}^n$, where $\{-1,0,1 \}^n$ denotes the set of $3^n$ possible vectors of length $n$ with coordinates in $\{-1,0,1 \}$. The $z$-scores for each feature are assumed to be independent across studies given the vector of hypotheses states, so the joint density of $(z_{i1}, \ldots, z_{in})$ for feature $i$ given $\boldsymbol{h}_i=(h_{i1},\ldots,h_{in})$ is $\prod_{j=1}^n f(z_{ij} \mid h_{ij})$.

The statistic that plays a central role for inference on the hypotheses states for each feature is the \emph{local false discovery rate}, which is the probability that the state vector is a null state vector given the feature's $z$-scores. For replicability analysis,   we wish to test the  $r/n$ no replicability null hypothesis  defined in (\ref{part-conj-dir1}), so the null state vector is a binary vector with at most $r-1$ positive entries and at most $r-1$ negative entries. We use the notation $\sum_{\boldsymbol{h}\in H _i^{r/n}}$ to denote the summation over all state vectors ${\boldsymbol{h}}$ that are in the null space of the $r/n$ no replicability null hypothesis.  
The local false discovery rate for feature $i$ is
\begin{eqnarray}
&& T_i = \sum_{{\boldsymbol{h}_0}\in H _i^{r/n}} \mP({\boldsymbol{h}_i}  = {\boldsymbol h_0} \mid z_{i1}, \ldots, z_{in}) 
\nonumber \\
&&= \frac{\sum_{{\boldsymbol{h}_0}\in H _i^{r/n}} \pi({\boldsymbol h_0})\prod_{j=1}^n f(z_{ij} \mid {h_0}_{j})}{\sum_{{\boldsymbol h}\in \{-1,0,1 \}^n} \pi({\boldsymbol h})\prod_{j=1}^n f(z_{ij} \mid h_{j})} \nonumber
\end{eqnarray}

With these local false discovery rates, we describe next  the steps for inference on the family of the $r/n$ no replicability null hypotheses, for a target $\alpha$ level  Bayes FDR. The Bayes  FDR is the probability that the hypothesis state vector is in the null space, given that the vector of $z$-scores is in the rejection region. For independent $z$-scores within each study, the Bayes FDR coincides with the positive FDR, which is the expected false discovery proportion given that at least one discovery was made \citep{Storey03}\footnote{The FDR is the positive FDR times the probability of making at least one discovery, so the FDR and positive FDR coincide  
in settings where always at least one discovery is made
\citep{Storey03}. 
}. The steps for inference are as developed for a single study in \cite{sun07}: 

\begin{algo}[\cite{heller2014replicability}]\label{algo-workflow-locfdr} $ $
\begin{steps}
    \item Compute for each $i \in \{1, \ldots, m\}$ the local false discovery rate $T_i$.
    \item Sort the $T_i$'s from smallest to largest, so $T_{(1)}\leq \ldots \leq T_{(m)}. $
    \item Reject the $r/n$ no replicability null hypothesis for the $R$ features with the smallest $T_i$'s, where 
    $$R = \max\left \lbrace k: \frac{\sum_{i=1}^k T_{(i)}}{k}\leq \alpha \right \rbrace. $$
\end{steps}
\end{algo}

In practice, the parameters  of the model are unknown. The empirical Bayes approach uses the data in order to estimate the parameters that are needed in order to compute the $T_i$'s.  
Methods for estimation of the mixture components, using the EM algorithm, have been suggested, e.g.,  in \cite{Chung14, heller2014replicability, Amar18}. 
These estimation methods  assume that the test statistics
(such as $p$-values) are independent in each study, which
is obviously not true for high dimensional studies. However, if the dependence is local (i.e., each feature's test statistic is associated only with a small number of other test statistics), the signal is not too sparse,  and the studies are high dimensional, the estimation method (which uses the composite likelihood rather than the true likelihood) is fairly robust to deviations from independence.  See  \cite{Xiang19} for another estimation method for the model parameters for $n=2$; See \cite{sunWei11, WangZhu19} and references within for methods that take the dependency between hypothesis states into account. 
\subsection{The (Bayes) FDR of a frequentist approach}
 The no-replicability null hypothesis is a composite hypothesis. The frequentist approach guards against the worst  (or least favorable) setting, while the empirical Bayes approach discussed in \S~\ref{subsec-eBayes} relies on the ability to estimate the prior probabilities for the different vectors of hypotheses states.

The methods in \S~\ref{sec:multiple studies} do not require  modeling of the $z$-values (or $p$-values). Moreover, they provide an  FDR (or FWER) control guarantee for any number of features $m$. This is in contrast with the inference based on the (estimated) local false discovery rate, which requires a model for the $z$-values, and can provide only asymptotic FDR control \citep{heller2014replicability}.

 To help clarify  the expected power advantage of the empirical Bayes approach of \S~\ref{subsec-eBayes} over the frequentist approaches in \S~\ref{sec:multiple studies}, when the model for the $z$-values approximates well the data generation mechanism, we provide the following analysis for the frequentist approach of applying BH on the PC $p$-values (although all procedures in \S~\ref{sec:frequentist} suffer from being conservative compared to the approach of \S~\ref{subsec-eBayes}, showing their conservativeness is more involved so it is outside the scope of this paper).
Consider a procedure that thresholds the PC $p$-values, so that only features with $p^{r/n}_i\leq t$ are rejected. The threshold $t$ may be selected to be the largest that satisfies $\frac{t}{\mP\left(p^{r/n}_i\leq t\right)}\leq\alpha$, where $$\mP\left(p^{r/n}_i\leq t\right) = \sum_{\boldsymbol{h}\in \{ -1,0,1\}^n}\mP\left(p^{r/n}_i\leq t\mid \boldsymbol{h}_i=\boldsymbol{h} \right)\pi(\boldsymbol{h})$$ is the marginal cumulative distribution function of the PC $p$-value.  Using this threshold provides a procedure that is a close approximation to the BH procedure on the PC $p$-values, which selects the largest $t$ so that  $\frac{t}{|\{i: p^{r/n}_i\leq t\}|/m} \leq \alpha$. The  positive FDR of this procedure for the family of $r/n$ no replicability nulls can be expressed in terms of any $i\in \{1,\ldots,m\}$ as follows:  
\begin{eqnarray}
&&\frac{\sum_{\boldsymbol{h}\in H_i^{r/n}}\mP\left(p^{r/n}_i\leq t\mid {\boldsymbol{h}_i=\boldsymbol{h}}\right)\pi(\boldsymbol{h})}{\mP\left(p^{r/n}_i\leq t\right)}\nonumber \\
&& \approx \frac{\sum_{\boldsymbol{h}\in H_i^{r/n}}\mP\left(p^{r/n}_i\leq t\mid {\boldsymbol{h}_i=\boldsymbol{h}} \right)\pi(\boldsymbol{h})}{|\{k: p^{r/n}_k\leq t\}|/m}.  \nonumber
\end{eqnarray}
The numerator is clearly at most $t$, but it may also be much smaller. For example, if the probability that the global null is true is 0.95 and that the no-replicability null is false is 0.05 (i.e., we can only observe features with $\boldsymbol{h} = \boldsymbol{0}$ or features with $\boldsymbol{h}\notin  H_i^{r/n}$),  then the numerator is $0.95\times\mP\left(p^{r/n}_i\leq t\mid {\boldsymbol{h}_i=\boldsymbol{0}}\right), $ which is much smaller than $t$ for $r\geq 2$ (e.g., for $r=2$ and $n=2$ the numerator is at most $0.95\times t^2$). Therefore, by choosing $t$ so that $\frac{t}{
|\{k: p^{r/n}_k\leq t\}|/m} \approx \alpha, $ the positive FDR is in fact much smaller than $\alpha$. On the other hand, with Procedure \ref{algo-workflow-locfdr}  the level is approximately $\alpha$ so it is not overly conservative and it may provide many more replicable findings. See \cite{heller2014replicability} for more comparisons of the empirical Bayes approach versus a frequentist approach for replicability. 

\subsection{Future directions}

The generalization of the two group model to $n>1$ studies examining the same $m$ features  is useful for the joint analysis of the $m$ features as long as the individual feature's mixture distribution can be estimated well. This is not possible if 
 the unknown dependence within each high dimensional study  is long range dependence (even when $m$ is in the order of $10^6-10^9$). It is not currently known how to develop a procedure for general (unknown) dependence structures within each study.  

\cite{Xie11, heller21} considered the following generalization of the two group model for   a single high dimensional study $j \in \{1,\ldots, n\}$ with known (or well approximated) dependence between the $z$-scores: the vector $(z_{1j}, \ldots, z_{mj})$ is sampled from the joint conditional distribution given $(h_{1j}, \ldots, h_{mj})$, where the $h_{ij}$'s are independently and identically distributed as in the two group model.   They showed that in order to discover the features with a false null hypothesis, the optimal statistic for each feature is the probability that the null hypothesis is true given the entire vector of $z$-scores, $\mP(h_{ij}=0 \mid z_{1j}, \ldots, z_{mj})$. The potential gain in power in using this statistic rather than the marginal statistic $\mP(h_{ij}=0 \mid z_{ij})$ can be very large. We thus expect the power advantage to be large when incorporating the within study local dependence in a joint analysis of $n$ studies. Specifically, for replicability analysis, it can be useful to incorporate known or approximate local dependence within each study, to boost discovery power of replicable signals. 

\cite{Xiang19} considered for $n=2$ several inferential goals, each with a predefined level of  asymptotic FDR control: classification of features to those that have signal in both studies, or only in the first study, or only in the second study. 
Their paper inspires additional interesting formulations for replicability analysis. For example, for $n=3$ studies,  consider simultaneously the classification of features with signal (in the same direction) in all three studies, as well as those with signal (in the same direction) in exactly two studies.

\section{The case of a single study}\label{sec:single study}
Splitting the sample at random into two halves is not useful for assessing replicability, since it results in having two sample, each with reduced sample size, for testing the same hypotheses $H_i, i=1, \ldots, m$.

Importantly,  splitting the sample according to an observed binary covariate is helpful for replicability. For example, examination of the effect of high fish consumption in a subgroup  that lives where fish are at low cost (near the coast) and in a subgroup that lives where fish are at high cost (say in midland, so the reason for consumption may be  health benefits)  may be possible from a publicly available database \citep{zhao2018cross}. If an outcome (e.g., heart disease) is negatively associated with increased fish consumption in both subgroups,  the evidence that the discovered association is due to fish consumption is more convincing and less susceptible to bias.
In this example,  the covariate indicates whether the subject lives near the sea or in midland, so it can be used to split the sample into two subgroups in which the effect of high fish consumption is examined. Although the same family of null hypotheses are considered in each subgroup (as in splitting at random), the hypothesis states need not be the same in the two resulting subgroups (unlike with random sample splitting). For example, it may be that there is no association between fish consumption and heart disease for the subgroup that lives near the sea, but there is a negative association for the subgroup that lives in midland  due to say unobserved confounding with exercise: in midland the reason for high fish consumption is health benefits, and those that are health conscious exercise more, and exercise reduces the risk of heart disease.   

When splitting according to an observed binary covariate, the aim is to identify features for which  the null hypothesis is false in both  subgroups, and the procedures applied should  take care not to declare as replicated too many features $i\in \{1,\ldots,m\}$ for which  the null hypothesis is false in exactly one of the subgroups.  The procedures in \S~\ref{sec: two studies} may therefore be appropriate for this purpose, see \cite{roy2023} for an example application. If the sample is split instead according to an observed covariate with $n\geq 2$ categories, the procedures in \S~\ref{sec:multiple studies} may be useful. 
 
The procedures in \S~\ref{sec: two studies} and \S~\ref{sec:multiple studies} may also be useful for causal inference in observational studies using evidence factors \citep{Karmakar20}. 
In order to establish causality in observational studies, it is useful to decompose the test of no treatment effect into evidence factors, which are separate pieces of information that are affected by different biases and are statistically nearly independent \citep{Rosenbaum10}. For example, in an observational study with a  matched pairs design, where each pair has one  exposed and one unexposed individual, and the level of exposure among the exposed is measured, then there are two pieces of evidence  towards the (say positive) effect of exposure: evidence that the exposed have higher levels of the outcome than their  unexposed matched pairs; evidence among the exposed of a positive association between exposure level and the outcome. See  \cite{Karmakar20} for examples with more than two evidence factors.  Suppose there are $n$ such pieces. \cite{Karmakar20} suggest testing whether there is evidence that at least $r$ of the $n$  pieces of information reject the null hypothesis of no treatment effect. This is the test of the no-replicability null hypothesis, and the PC $p$-value is used to evaluate this composite null.

\section{Discussion}
\cite{Wang20} point out that  when  complicated large-scale experiments or observational studies are carried out, they can be susceptible to many sources of bias. 
  Since many complicated large-scale studies are  carried out nowadays, it is crucial to have statistical methods that objectively evaluate the consistency of findings across the studies, while properly accounting for the fact that the parameters of interest may differ  across studies. More specifically, the consistency (or replicability) across studies has to be evaluated by ruling out that the finding is particular to a single study (or more generally, to $r-1$ studies). In this review, we provided  various relevant procedures that may be useful. 

In collaborative research, it is possible to go one step further in terms of the flexibility of the procedures towards establishing replicability. \cite{roy2023} suggests for $n=2$ that two separate teams will engage in the design of the multiple testing, where each team has access only to a single study, and by  examining that study it designs the multiple testing of the other study. As long as the teams act independently, they have complete freedom in design decisions. So a more flexible version of Procedure \ref{Biometrika-FWER} will allow each team not only to select hypotheses and their direction of testing, but also to select  the test statistic for each hypothesis,  possible weights for the different hypotheses (which may reflect the prior belief in the probability that they are nonnull in the other study), and the FWER controlling procedure to apply in the other study. For $n>2$, each team may have access to a group of studies, and the additional flexibility of choosing how to combine the evidence from the group of studies. If there are more than two teams, the flexibility is further increased. Procedure \ref{Biometrika-FWER} has to  be generalized, and we expect it to be a powerful and useful tool for tailoring replicability analyses to specific applications. 

The suggested procedures in this review were  motivated by the need for replicable results that are less susceptible to biases. However, these methods  are also  useful in additional settings. For example, in mediation analysis \citep{Sampson18, Vera19, Liu22}, where the aim is to identify the mediators that are associated both with the exposure and with the outcome. As another example, in a group analysis of size $n$, where multiple features are examined for each subject, and the goal is to find the features with signal in at least $r/n$ subjects. Specifically, consider the  neuroimaging application addressed by \cite{Heller07}, where the aim is to identify the brain regions that are consistently activated in a group of subjects.  

Increasingly, researchers provide access to the data used in their studies, at least in aggregate form. For example, in GWAS, summary statistics for each genotype are available in various consortia, and $p$-values can be easily computed from the summary level data. So the suggested $p$-value based replicability analyses can include studies from these consortia. We hope the facilitation of access to data will increase the practice of applying a replicability analysis to establish scientific findings.

\section*{Acknowledgment}
The first author was supported in part  by Israeli Science Foundation grant no. 1726/19.
The second author was supported in part  by Israeli Science Foundation grant no. 2180/20.
 The authors are  grateful to Yoav Benjamini, Saharon Rosset, and Daniel Yekutieli  for useful discussions that helped shape the paper.


\appendix

\section{Supplementary examples}\label{sec-SM}
This section provides several additional examples which led to the conclusions presented in Section 5.1 of the main manuscript. The examples address the naive approach considered in Section 5.1 of the main manuscript, where  the analyst declares as replicated the features in the set
$\mathcal{D}_1\cap (\cup_{i=2}^n \mathcal{D}_i),$ where $\mathcal{D}_1$ is the set of features discovered in the given study, called study one, and $\mathcal{D}_i,\,i=2, \ldots, n$ are the sets of features discovered in $n-1$ previous studies, where each $\mathcal{D}_j$ is the result of a certain multiple testing procedure.
\begin{example}\label{ex: small-FWER}
Consider the case where the given study examines $m=2$ features, and the researchers assess replicability based on one previous study examining the same two features.
Assume that feature 1 has a strong effect in study one. So in this idealized setting, we assume $p_{11}= 0$ with probability 1, and  no effect in study 2, which gives $p_{12}\sim U(0,1).$ The configuration for feature 2 is opposite with respect to the studies: it has a strong effect in study two and no effect in study 1, which gives $p_{21}\sim U(0,1),$ and $p_{22}\approx 0$ with probability 1.
See Table \ref{tab:FWER-small} for illustration.
\begin{table}[h]
\centering
	\begin{tabular}{|*{6}{c|}}
		\hline
			& Study 1 & Study 2  \\\hline
			 1 & 0 & $U(0,1)$ \\\hline
			2 & $U(0,1)$ & 0  \\\hline
				\end{tabular}
		\caption{Illustration of the setting addressed in Example \ref{ex: small-FWER}. The $ij$'th entry  indicates the distribution of $p_{ij},$ for each $i,j\in\{1,2\}:$ $0$ indicates that the given $p$-value is approximately 0 with probability 1, $U(0,1)$ indicates that its distribution is uniform.}
		\label{tab:FWER-small}
\end{table}
In this case any replicability claim is false, because each feature has an effect only in one study. Assume one applies Holm's or Hochberg's procedure at level $\alpha$ within each study, targeting FWER control. For each of these procedures,  $H_{11}$ and $H_{22}$ are necessarily rejected, and each of the other hypotheses is rejected with probability $\alpha.$   In this setting, rejection of $H_{12}$ leads to a false replicability claim for feature 1, and rejection of $H_{21}$ leads to a false replicability claim for feature 2. Since the studies are independent, we obtain that the probability of making one or more false replicability claims, which is the FWER for replicability analysis, is $1-(1-\alpha)^2>\alpha.$

\end{example}
In practice $n$ may be much larger than two. In this case, when $m$ is small, the inflation of the FWER for replicability analysis may be more significant. The following example illustrates this.
\begin{example}\label{ex:FWER-medium}
Consider the setting where two features are examined in $n>2$ studies. In study one and two, the configuration is the same as in Example \ref{ex: small-FWER}. In the other studies, both features have no effects and have uniform $p$-values. 
See Table \ref{tab:FWER-medium} for illustration.
\begin{table}[h]
  \centering
  	\begin{tabular}{|*{6}{c|}}
			\hline
			& Study 1 &Study 2 &  Study 3 &$\ldots$ & Study $n$  \\\hline
			 1 & 0 & $U(0,1)$ & $U(0,1)$ & \ldots & $U(0,1)$ \\\hline
			2 & $U(0,1)$ &  0 & $U(0,1)$ & $\ldots$ &  $U(0,1)$ \\\hline
					\end{tabular}
   \caption{Illustration of the setting addressed in Example \ref{ex:FWER-medium}. The $ij$'th entry  indicates the distribution of $p_{ij},$ for each $i\in\{1, 2\},$ $j\in\{1,\ldots, n\}:$ $0$ indicates that the given $p$-value is approximately 0 with probability 1, $U(0,1)$ indicates that its distribution is uniform.}
   \label{tab:FWER-medium}
   \end{table}
In this case, as in Example \ref{ex: small-FWER}, any replicability claim is false. Assume that each study is tested using Holm's procedure at level $\alpha.$ 
Since $H_{11}$ and $H_{22}$ are necessarily rejected, minimal replicability  is claimed for feature 1 if it is discovered in at least one study among studies $2, \ldots, n,$ and it is claimed for feature 2 if it is (erroneously) discovered in study one. Hence, the probability of making a false replicability claim for feature 1 is higher than in the setting of Example \ref{ex: small-FWER}, which leads to a higher FWER for replicability analysis. In order to obtain its value, let us first compute the probability of rejecting at least one hypothesis in the set $\{H_{13}, \ldots, H_{1n}\}.$ It is clear that in this setting, the probability of rejecting $H_{1j}$ is the same for $j\in\{3, \ldots, n\}.$ For a fixed $j$ in this set,  it is equal to $f(\alpha)\equiv \mP(p_{1j}\leq \alpha/2)+\mP(\alpha/2<p_{1j}\leq\alpha, p_{2j}\leq \alpha/2).$ Under independence within each study, $f(\alpha)=\alpha/2+\alpha^2/4.$ Therefore, the probability of making at least one false replicability claim for feature 1 is $1-(1-\alpha)(1-f(\alpha))^{n-2}.$ 
The FWER for replicability analysis is $1-(1-\alpha)^2(1-f(\alpha))^{n-2},$ and it approaches 1 as $n$ increases. For $n=10$ and $\alpha=0.05,$ under independence within each study, this probability is  $1-(0.95)^2(1-0.0256)^8\approx0.27.$ 
\end{example}
\begin{example}\label{ex:FWER-big2}
Consider the setting where $m>3$ features are examined in $n>2$ studies. For features 1 and 2, the configuration is the same as in Example \ref{ex:FWER-medium}. Feature 3 has false null hypotheses in all the studies, with $p_{3j}=0$ with probability 1 for each $j\in\{1, \ldots, n\}.$ The remaining features, with indices $i\in\{4, \ldots, m\},$ have true null hypotheses in all the studies, with $p_{ij}\sim U(0,1)$ for $j=1, \ldots, n.$ See Table \ref{tab:FWER-big2} for illustration.
\begin{table}[h]
   \centering
   	\begin{tabular}{|*{6}{c|}}
			\hline
			& Study 1 &Study 2 &  Study 3 &$\ldots$ & Study $n$  \\\hline
			 1 & 0 & $U(0,1)$ & $U(0,1)$ & \ldots & $U(0,1)$ \\\hline
			2 & $U(0,1)$ &  0 & $U(0,1)$ & $\ldots$ &  $U(0,1)$ \\\hline
				3 & 0 &  0 & 0 & $\ldots$ &  0\\\hline
				4 & $U(0,1)$ & $U(0,1)$ & $U(0,1)$& $\ldots$ & $U(0,1)$ \\\hline
				$\ldots$ &$\ldots$&$\ldots$&$\ldots$&$\ldots$ &$\ldots$ \\\hline
				$m$ & $U(0,1)$ & $U(0,1)$ & $U(0,1)$ & $\ldots$ &$U(0,1)$\\\hline
		\end{tabular}
    \caption{Illustration of the setting addressed in Example \ref{ex:FWER-big2}. The $ij$'th entry indicates the distribution of $p_{ij},$ for each $i\in\{1, \ldots, m\},$ $j\in\{1,\ldots, n\}:$ $0$ indicates that the given $p$-value is 0 with probability 1, $U(0,1)$ indicates that its distribution is uniform.}
    \label{tab:FWER-big2}
\end{table}
 Assume that the discoveries within each study are obtained using the level-$\alpha$ Bonferroni procedure. In this case, the probability of making  a false minimal replicability claim for each of the features is  given in Table \ref{tab:FWER-big-prob}.
\begin{table}[h]
   \centering
   	\begin{tabular}{|*{2}{c|}}
			\hline
		Feature	& $\mP(\text{false minimal replicability claim})$  \\\hline
			 1 & $g_1(\alpha)=1-(1-\alpha/m)^{n-1}$ \\\hline
			2 & $g_2(\alpha)= \alpha/m$ \\\hline
				$i\in\{4, \ldots, m\}$ & $g_3(\alpha)=(\alpha/m)\left[1-(1-\alpha/m)^{n-1}\right]$  \\\hline
		\end{tabular}
    \caption{The probability of making a false replicability claim for features with indices in the set $\{1,2, 4,5, \ldots, m\}$ under the setting of Example \ref{ex:FWER-big2}. The probability of making a false replicability claim for any feature with index $i\in\{4, \ldots, m\}$ is the same, and its common value is given in the last row.} 
    \label{tab:FWER-big-prob}
\end{table}
Therefore, under independence within each study, the FWER for replicability analysis is equal to:
\begin{align*}&\mP(\text{at least one false replicability claim})=\\&1-\left[1-g_1(\alpha)\right]\left[1-g_2(\alpha)\right]\left[1-g_3(\alpha)\right]^{m-3}=\\&1-(1-\alpha/m)^{n}\left\{1-(\alpha/m)\left[1-(1-\alpha/m)^{n-1}\right]\right\}^{m-3}. \end{align*}
When $m$ is large and $n<<m,$ so that $n/m$ is close to 0, the FWER for replicability analysis is close to 0. 
For $\alpha=0.05,$ and  $n=10$ the FWER for replicability analysis is: $5\times 10^{-7}$ for $m=10^6$;  0.0052 for  $m=100$.  When $m$ is large and $n\approx m,$ the FWER for replicability analysis is approximately $\alpha(1+\alpha(1-\exp\{-\alpha\}),$ which  is only slightly larger than $\alpha$ for small values of $\alpha.$ For $\alpha\in\{0.05, 0.1\},$  the approximated FWER values are 0.0511, 0.1037, respectively. The approximated values are close to the exact ones even for rather small $m$: for $\alpha=0.05,$ $m=n=50,$ the exact value of FWER is 0.0509, and for  $m=n=100,$ it is 0.0510. For small $m$ and $n>>m,$ the FWER for replicability analysis may approach 1. For $\alpha=0.05,$ $m=10$ and $n=100,$ the FWER is 0.4. 
\end{example}
\begin{example}\label{ex:big:gen}
Consider the generalized setting  of Example \ref{ex:FWER-big2}, where $m$ features (where $m$ is large) are examined in $n>3$ studies, and the features have four possible configurations: configuration of feature 1, feature 2, feature 3, and feature 4 in Table \ref{tab:FWER-big2}. Let $\pi_i$ be the proportion of features with configuration of feature $i$, for $i=1,2,3,4,$  so $\pi_1+\pi_2+\pi_3+\pi_4=1.$ Assume that the $p$-values within each study are independent. Then we obtain: 
\begin{align*}
  &\mP(\text{no false minimal replicability claims})=\\&(1-\alpha/m)^{(n-1)m\pi_1}\times(1-\alpha/m)^{m\pi_2}\times\\&\left\{1-(\alpha/m)[1-(1-\alpha/m)^{n-1}]\right\}^{m\pi_4}.
\end{align*}
In the setting of Example \ref{ex:FWER-big2}, $\pi_1=\pi_2=\pi_3=1/m,$ while $\pi_4=(m-3)/m,$ so for large $m$ these proportions are either close to 0 or close to 1. Let us now assume that $m$ is large, $n<<m$ so that $n/m$ is close to 0, and $\pi_1,$  the proportion of features with an effect in study one which is non-replicable in other studies, is fixed and positive. In this setting, the expression above can be approximated by $\exp\{-\alpha[(n-1)\pi_1+\pi_2],$ so the FWER for replicability analysis is approximately $1-\exp\{-\alpha[(n-1)\pi_1+\pi_2]\}.$ If $(n-1)\pi_1+\pi_2\leq1,$ then the approximated value of FWER is at most $1-\exp\{-\alpha\}\approx \alpha,$ for small $\alpha.$ For example, for $n=10,$ $\pi_1=\pi_2=0.1,$ $(n-1)\pi_1+\pi_2=1,$ and $\text{FWER}\approx\alpha.$ However, if  $(n-1)\pi_1+\pi_2>1,$ then the approximated value of FWER may be far higher than $\alpha.$ 
For example, for $n=10,$ $\pi_1=\pi_2=0.2,$ and $\alpha=0.05,$ the FWER is approximately 0.095, which is almost twice the nominal level.  For $\pi_1=0.5,$ $\pi_2=0,$ $n=10,$ and $\alpha=0.05,$ the FWER is approximately 0.2. The real values of FWER for the above configurations are 
the same as the approximated ones up to three digits after the decimal point even for $m=100,$ and the approximation is even more precise for larger $m.$ Obviously, in this configuration, increasing at least one of the parameters in the set $n,\pi_1, \pi_2,$ while keeping the others fixed, increases the FWER for replicability analysis. When  $m$ is large and $n\approx m,$ while $\pi_1$ is fixed and positive, the FWER for replicability analysis may be close to 1. 
For example, for $m=n=100,$ $\alpha=0.05,$ $\pi_1=\pi_2=0.1,$ $\pi_4=0.7,$ the value of FWER is  0.3946. Setting $\pi_1=0.5,$ $\pi_2=0,$ $\pi_4=0.4$ while keeping the other parameters the same, we obtain FWER of 0.916. Increasing $m,n$ to 500 gives $\text{FWER}=0.918$ and $\text{FWER}=0.999$ for the first set and second set of parameters, respectively.
\end{example}
The examples above address the idealized setting where for each false null hypothesis, the $p$-value is equal to 0 with probability 1. The following example addresses more realistic distributions of the $p$-values for false null hypotheses. 
\begin{example}
 Consider the setting of Example \ref{ex:FWER-big2} with a modified configuration:  the cumulative distribution function (CDF) of the $p$-values corresponding to false null hypotheses (indicated by 0 in Table \ref{tab:FWER-big2}) is given by $F,$ which is different from that of Table \ref{tab:FWER-big2},  while the distribution of $p$-values corresponding to true null hypotheses remains $U(0,1).$   Then the probability of making a false replicability claim for feature 1 is smaller than in Example \ref{ex:FWER-big2}. Specifically, it is equal to $F(\alpha/m)g_1(\alpha),$  where $g_1(\alpha)$ is the probability of this event under the configuration of Table \ref{tab:FWER-big2}, and it is given in the first row of Table \ref{tab:FWER-big-prob}. Similarly, the probability of this event for feature 2 is $g_2(\alpha)\{1-[(1-F(\alpha/m))(1-\alpha/m)^{n-2}]\},$ where $g_2(\alpha)$ is the probability of this event under the configuration of Table \ref{tab:FWER-big2}, and it is given in the second row of Table \ref{tab:FWER-big-prob}. Clearly, the probability of making a false replicability claim for feature $i\in\{4, \ldots, m\}$ is the same as in Example \ref{ex:FWER-big2}, i.e. it is  $g_3(\alpha),$ given in the third row of Table \ref{tab:FWER-big2}.
 Therefore, under independence within each study, the FWER for replicability analysis under the current configuration is smaller than under the configuration of Table \ref{tab:FWER-big2}. Assume that any pair $(i,j),$ $p_{ij}$ is computed based on a normal test statistic whose distribution is $N(0,1)$ if $H_{ij}$ is true, and is $N(\mu, 1)$ if $H_{ij}$ is false,  where $\mu>0.$ Then $F(\alpha/m)=1-\Phi(\Phi^{-1}(1-\alpha/m)-\mu),$ where $\Phi$ is the CDF of the standard normal distribution, and $\Phi^{-1}$ is its inverse function.
 Let us consider the sets of parameters for which we observed a considerable inflation of FWER under the settings of Examples \ref{ex:FWER-big2} and \ref{ex:big:gen}. For all these sets, $\alpha=0.05.$ Under the setting of Example \ref{ex:FWER-big2}, for  $m=10,$ $n=100,$ and $\mu=2,3$ the FWER is 0.125 and 0.273, respectively (for large $\mu,$ which approximately corresponds to the case of $p$-value equal to zero with probability 1, $\text{FWER}\approx 0.4$). Under the setting of Example \ref{ex:big:gen}, for $n=10,$ $\pi_1=\pi_2=0.2,$ $\pi_3=0.1,$ $\pi_4=0.5$ and $m=500,$ FWER is 0.023 and 0.059 for $\mu=3$ and $\mu=4,$ respectively (its value is approximately 0.095 for large $\mu$).  For $\pi_1=\pi_4=0.5,$ $n=10,$ and $m=500,$ FWER is 0.052 and 0.128 for $\mu=3$ and $\mu=4,$ respectively (its value is approximately 0.2 for large $\mu$). When the number of studies is the same as the number of features, the inflation of FWER becomes more severe: for $m=n=100,$ $\pi_1=\pi_2=\pi_3=0.1,$ $\pi_4=0.7,$  FWER is 0.175 and 0.316 for $\mu=3$ and $\mu=4$ respectively (its value is approximately 0.395 for large $\mu$). Setting $\pi_1=0.5, \pi_2=0, \pi_3=0.1, \pi_4=0.4$ while keeping $m$ and $n$ fixed to 100, we obtain FWER of 0.21, 0.61 and 0.85 for $\mu=2,3,4,$ respectively. Increasing $m,n$ to 500 gives FWER values of 0.44 and 0.94 for $\mu=3,$ and   $(\pi_1, \pi_2, \pi_3, \pi_4)$ set to $(0.1, 0.1, 0.1, 0.7)$ and  $(0.5, 0, 0.1, 0.4),$ respectively. For $\mu=4,$ the FWER values for these sets of parameters increase to 0.78 and 0.99, respectively.
\end{example}

\begin{example}
Consider the setting where $m$ features are examined in $n$ studies, and all the features have no effects in all the studies, leading to uniform $p$-values. Assume that the discoveries within each study are obtained using level-$\alpha$ Bonferroni procedure. 
 Under independence within each study, we obtain:
\begin{align*}
   &\mP(\text{at least one false minimal replicability claim})=\\&
    1-\left\{1-(\alpha/m)[1-(1-\alpha/m)^{n-1}]\right\}^{m}
\end{align*}
In the setting where $m$ is large and $n<<m,$ so that $n/m$ is close to 0, this expression is close to 0. 
In the setting where $m$ is large and $n\approx m,$ this expression is close to $1-\exp\{-\alpha(1-\exp\{-\alpha\})\},$ which is approximately 
$\alpha^2$ for small $\alpha.$ 
For large $n$ and $m<<n,$ so that $m/n$ is close to 0, the expression above is approximately $1-(1-\alpha/m)^m,$ which is upper bounded by $\alpha.$ 
Therefore, in this setting, the naive approach is valid in all the above scenarios. However, when an investigator faces $n$ symmetric studies, and minimal replicability claims are made for features which have discoveries in any two of the $n$ studies (i.e. whose indices belong to $\cup_{i\neq j}\mathcal{D}_i\cap \mathcal{D}_j$), 
then the situation is different. The probability of having at least two common discoveries for at least one feature is
\begin{align*}
    1-[(1-\alpha/m)^n+n(1-\alpha/m)^{n-1}\alpha/m],\label{FWER-symmetric}
\end{align*}
so its value is close to 1 when $n$ is large and $m<<n.$ For $\alpha=0.05,$ $m=5$ and $n=100,$ its value is 0.264.
\end{example}

\end{document}